\documentclass{article}
\usepackage{amsmath,amssymb,amsbsy,graphics,graphicx,bbm}
\usepackage[a4paper,portrait]{geometry}
\usepackage{setspace}
\usepackage{color}
\usepackage{ulem}
\usepackage[pdftex,dvipsnames]{xcolor}
\usepackage[linecolor=OliveGreen,backgroundcolor=OliveGreen!25,bordercolor=OliveGreen,textsize=small]{todonotes}
\newtheorem{prop}{Proposition}
\newtheorem{thm}{Theorem}
\newtheorem{Remark}{Remark}
\newtheorem{assm}{Assumption}

\newtheorem{lemm}{Lemma}
\newtheorem{corr}{Corollary}

\renewcommand{\P}{\mathbb{P}}
\newcommand{\R}{\mathbb{R}}
\newcommand{\N}{\mathbb{N}}

\newcommand{\E}{\mathbb{E}}

\begin{document}

\doublespace
\title{On the sign recovery  by LASSO, thresholded LASSO and  thresholded Basis Pursuit Denoising}
\author{ Patrick J.C. Tardivel$^{a}\footnote{Corresponding author: tardivel@math.uni.wroc.pl}\,$ and 
Ma\l{}gorzata Bogdan$^{a,b}$,\\
\small{ $^{a}$  Institute of Mathematics, University of Wroc{\l}aw, Wroc\l{}aw, Poland }\\
\small{ $^{b}$  Department of Statistics, Lund University, Lund, Sweden}}
\date{}
\maketitle

\begin{abstract}
    Basis Pursuit (BP), Basis Pursuit DeNoising (BPDN), and LASSO are popular methods for identifying important predictors in the high-dimensional linear regression model $Y=X\beta+\varepsilon$. 
By definition, when $\varepsilon=0$, BP uniquely recovers $\beta$ when $X\beta=Xb$ and $\beta\neq b$ implies $\|b\|_1>\|\beta\|_1$ (identifiability condition). Furthermore, LASSO can recover the sign of $\beta$ only under a much stronger irrepresentability condition. Meanwhile, 
it is  known that the model selection properties of LASSO can be improved by 
hard-thresholding its estimates. This article supports these findings by proving that thresholded LASSO, thresholded BPDN and thresholded BP recover the sign of $\beta$ in both the noisy and noiseless cases if and only if $\beta$ is identifiable and large enough. In particular, if $X$ has iid Gaussian entries and the number of predictors grows linearly with the sample size, then these thresholded estimators can recover the sign of
$\beta$ when the signal sparsity is asymptotically below the Donoho-Tanner transition curve. This is in contrast to the regular LASSO, which asymptotically, recovers the sign of $\beta$ only when the signal sparsity tends to 0. Numerical experiments show that the identifiability condition, unlike the irrepresentability condition, does not seem to be affected by the structure of the correlations in the $X$ matrix.
\end{abstract}


\section{Introduction}
Let us consider the high dimensional linear model
\begin{equation} \label{model0} Y=X\beta+\varepsilon,\end{equation}
where $X=(X_1|\dots|X_p)$ is a $n \times p$ design matrix with $n \le p$, $\varepsilon$ is a random vector in $\R^n$, and $\beta\in\R^p$ is an unknown vector of regression coefficients. 
The sign vector of $\beta$ is
$S(\beta)=(S(\beta_1),\dots S(\beta_p))\in \{-1,0,1\}^p$, 
where for $x\in \R$, $S(x)={\bf 1}_{x > 0}-{\bf 1}_{x< 0}$.
Our main goal is to recover $S(\beta)$. This goal is somewhat more general than the goal of recovering the active set, 
${\rm supp}(\beta):=\{i \in \{1,\dots,p\} \mid \beta_i \neq 0\}$. Indeed, given $S(\beta)$ one may  recover ${\rm supp}(\beta)$ since 
${\rm supp}(\beta):=\{i \in \{1,\dots,p\} \mid S(\beta_i) \neq 0\}$ but conversely, given  ${\rm supp}(\beta)$, one cannot recover $S(\beta)$. Moreover, some theoretical works recommend sign recovery instead of active set recovery \cite{gelman2000type,stephens2017false}.

\subsection{BP, BPDN and LASSO}

First introduced in signal processing \cite{chen1994basis}, BP estimator is a solution of the 
 following optimization problem 
$$
\hat\beta^{\rm BP}:={\text{\rm argmin }}\|b\|_1\;\; \text{\rm subject to } Y = Xb,
$$
which can be solved using  linear programming.
In the noiseless case, when $\varepsilon=0$ and $Y=X\beta$,   we say that 
$\beta$ is identifiable with respect to $X$ and the $\ell_1$ norm if $\hat \beta^{\rm BP}$ uniquely recovers $\beta$  
{\it i.e.} if $b\neq \beta$, $Xb=X\beta$ implies $\|b\|_1>\|\beta\|_1$. Note that
a geometric characterization of the identifiability condition, depending on $X$ and the sign of $\beta$, is given in
\cite{schneider2020geometry}. 
 This condition implies that
$\beta$ is sparse. Indeed, Lemma 3 in Tardivel et al.  \cite{tardivel2018sparsest}
shows that ${\rm card}\{i \in \{1,\dots,p\}\mid \hat \beta^{\rm BP}_i\neq 0\}\le n$ i.e. $\hat \beta^{\rm BP}$ has at least $p-n$ zeros. Consequently having $k\le n$,  where $k$ is the number of non-zero elements in $\beta$, is  necessary for the  identifiability condition.  
 On the other hand, some assumptions on the sparsity of $\beta$ guarantee that $\beta$ is identifiable with respect to the $\ell_1$ norm. For example, the identifiability condition holds if 
 the number of non-zero elements of $\beta$ satisfies the mutual coherence condition \cite{donoho2003,foucartlivre,gribonval}.
In the special case where the elements of $X$ are iid normal variables, the transition curve $\rho:(0,1)\mapsto (0,1)$ \cite{DonTan05} characterizes the 
identifiability condition with respect to the signal sparsity $\xi=k/n$. In particular, it is known that the {\it identifiability} condition is satisfied with probability converging to 1 if $\xi < \rho(\delta)$ or converging to 0 if $\xi > \rho(\delta)$.

In the case where $\varepsilon$ has a continuous distribution, almost surely BP has $n$ non-zero components and thus BP cannot recover  the sign of $\beta$ once $k\neq n$. To account for noise, when $\varepsilon$ has iid $\mathcal{N}(0,\sigma^2)$ entries for some $\sigma>0$, Chen and Donoho \cite{chen1994basis} extended BP by proposing the following estimation algorithm: \begin{equation}\label{LASSOpen}\widehat{\beta}^{\rm L}:=\underset{b\in \R^p}{\text{\rm argmin }}\frac{1}{2}\|Y-Xb\|_2^2+\lambda\|b\|_1\text{ \rm with }\lambda=\sigma \sqrt{2\log p}.\end{equation}
In the seminal work of Tibshirani \cite{tibshirani1996}, the algorithm was extended for fitting the general multiple regression models where different values of $\lambda$ are often more appropriate. The method gained great popularity in the statistical community under the name LASSO (Least Absolute Shrinkage and Selection Operator), while in the signal processing community it is often called Basis Pursuit Denoising (BPDN).
In this paper, we will use the term BPDN for a slightly different form of this estimator, where association with BP is even clearer 
\begin{equation}\label{BPDN}
\widehat{\beta}^{\rm BPDN}:=\underset{b\in \R^p}{\text{\rm argmin }}\|b\|_1 \;\text{\rm subject to }\|Y-Xb\|_2^2\le R, \text{\rm where }
R\ge 0.
\end{equation}
Note that $X$ is fixed but $Y$ is random, and that for a given $\lambda> 0$ for LASSO we cannot choose a fixed $R >0$ for BPDN under which the two estimators are almost surely equal. 
 To illustrate this fact, we  note that $\widehat \beta^{\rm BPDN}=0$ if and only if $\|Y\|_2^2\le R$, while for LASSO $\widehat \beta^{\rm L}=0$ if and only if $\|X'Y\|_\infty\le \lambda$.

\subsection{Sign recovery by LASSO}

Properties of the LASSO sign estimator $S(\widehat{\beta}^{\rm L}(\lambda)):=\left(S(\widehat{\beta}^{\rm L}_1(\lambda)),\dots,
S(\widehat{\beta}^{\rm L}_p(\lambda))\right)$ (or properties of the active set estimator 
${\rm supp}(\widehat{\beta}^{\rm L}(\lambda)):=\{i \in \{1,\dots,p\} \mid \widehat{\beta}^{\rm L}_i(\lambda)\neq 0\}$)
	have been intensively studied \cite{dossal2012,lounici,meinshausenhigh,wainwright2009,zhao,zou}. 
According to Theorem 2 of Wainwright \cite{wainwright2009}, the irrepresentability condition is necessary for LASSO to recover $S(\beta)$ with high probability. Indeed, if $\ker(X_I)=\{{\bf 0}\}$, 
$\|X_{\overline{I}}'X_I(X_I'X_I)^{-1}S(\beta_I)\|_\infty> 1$ and both $\varepsilon$ and $-\varepsilon$ have the same distribution, then for any choice of the tuning parameter $\lambda> 0$, 
 $\P(S(\widehat{\beta}^{\rm L}(\lambda))=S(\beta))\le 1/2$. This result also holds for the noiseless case, in which LASSO does not recover $S(\beta)$.
Moreover, B\"{u}hlmann and van de Geer \cite{buhlmannlivre} (pp. 192-194) have
shown that if $\varepsilon={\bf 0}$ and the irrepresentability holds strictly (i.e., if 
$\|X_{\overline{I}}'X_I(X_I'X_I)^{-1}S(\beta_I)\|_\infty < 1$) then the nonrandom set
${\rm supp}(\widehat \beta^L(\lambda))$ is equal to ${\rm supp}(\beta)$ as long as the non-zero components of $\beta$ are sufficiently large. The proof given in \cite{buhlmannlivre} can be easily adapted for sign recovery.

Proposition \ref{CI_implique_NSP}, proved in the Appendix, shows that the identifiability condition is weaker than the 
irrepresentability condition.
\begin{prop}
\label{CI_implique_NSP}
Let $X$ be an $n\times p$ matrix with $p\geq n$ columns in general position. Moreover, let $\beta\in \R^p$, $I:={\rm supp}(\beta)$ and suppose that $\ker(X_I)=\{{\bf 0}\}$. 
If the irrepresentability condition holds, then the parameter $\beta$ is identifiable with respect to the $\ell_1$norm.
\end{prop}
 
For the case where the elements of $X$ are iid random variables from the normal distribution, the irrepresentability condition is satisfied with high probability if and only if $k< \frac{n}{2\log p}$ (see \cite{wainwright2009}). This implies that LASSO cannot recover $S(\beta)$ when $k/n\to \xi > 0$, even if $\xi <\rho(\delta)$.

\subsection{Sign recovery by thresholded LASSO}

It is well known that LASSO can consistently estimate $\beta$ under much weaker assumptions than the irrepresentability condition (see, {\it e.g.}, \cite{meinshausenlasso} or \cite{vandegeer2009}).  This suggests that an appropriately thresholded version of LASSO can recover $S(\beta)$ under weaker assumptions than the irrepresentability condition \cite{pokarowski2019improving}.
Concerning sign recovery by thresholded BP, first theoretical results were given by Saligrama and Zhao \cite{saligrama}. More recently, Descloux and Sardy \cite{descloux} 
show that the stable null space property is a sufficient condition to recover the sign of $\beta$ by thresholded BP. 

To our knowledge, the stable null space property has so far been the weakest sufficient condition known to guarantee the recovery of $S(\beta)$. 
In Theorem \ref{convergence}, we show that ``identifiability'', a weaker condition than the stable null space property,  is sufficient to recover $S(\beta)$ by thresholded BP, thresholded BPDN or thresholded LASSO when the non-zero elements of $\beta$ are large enough.
We also show that ``identifiability'' is a necessary condition for the recovery of $S(\beta)$ by these methods, regardless of the signal %size
magnitude.

Theorem \ref{convergence} implies that in the linear sparsity regime of \cite{DonTan05} for Gaussian matrices, thresholded BP, thresholded BPDN and thresholded LASSO  recover $S(\beta)$ if, asymptotically, $k<n\rho(n/p)$ holds and the signal %size 
magnitude tends towards infinity. 
Table \ref{tab:my_label} summarizes known conditions for sign or support recovery by LASSO and related methods, when $\varepsilon\neq 0$ and the signal is large enough.
\begin{table}[h]
    \centering
    \begin{tabular}{c|c|c|c}
      Method & Conditions    & Type & 
      Related articles 
 \\
 \hline \
Regular & Mutual coherence    & sufficient
 &  \cite{candesplan,lounici}\\
 LASSO &Irrepresentability  & necessary and sufficient 
 & \cite{dossal2012,meinshausenhigh,wainwright2009,zhao,zou}\\
 \hline
 Thresholded&Stable null space & sufficient
 & \cite{descloux}\\
 LASSO, BP&Identifiability&necessary and sufficient&current article
    \end{tabular}
    \caption{Some theoretical conditions under which LASSO and  related methods can recover the sign or the support of $\beta$ when $\varepsilon\neq 0$ and non-zero elements of $\beta$ are large enough. Additional comments related to these conditions  are given in Appendix.}
      \label{tab:my_label}
\end{table}

\subsection{Graphical illustration of the main result}

\begin{figure}[htbp]
\begin{tabular}{c c}
	\includegraphics[scale=0.5]{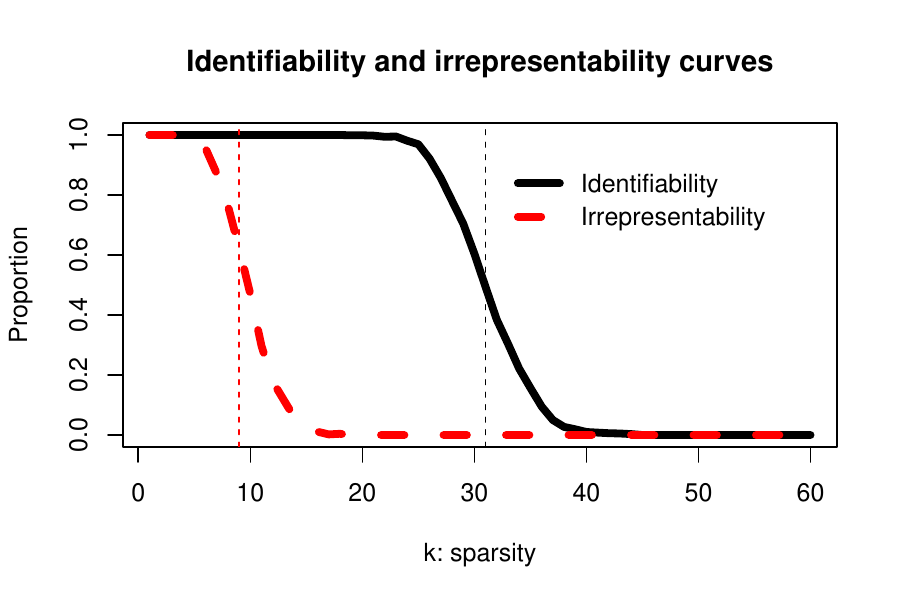}&
		\includegraphics[scale=0.5]{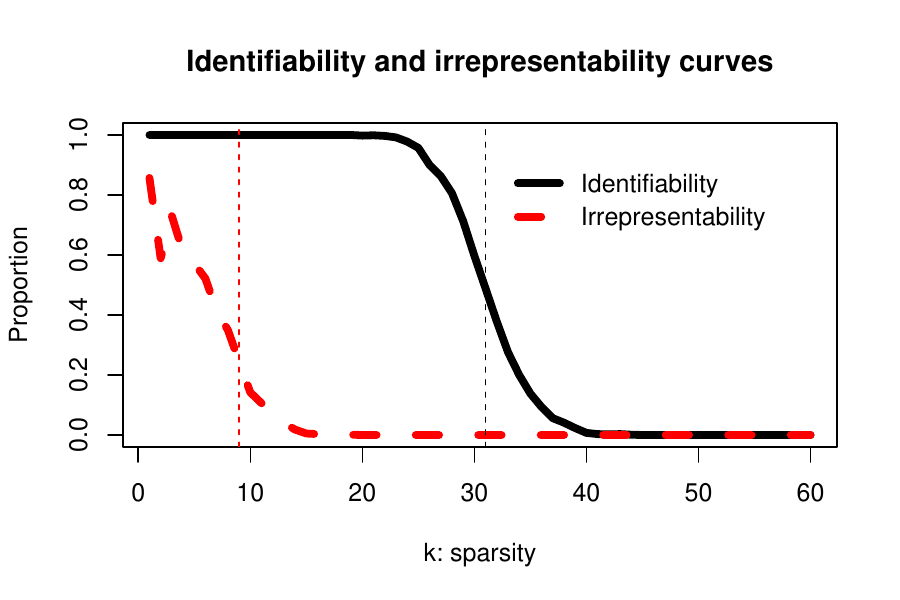}
		\end{tabular}
	\caption{Simulated identifiability and 
 irrepresentability curves for two specific design matrices of dimensions $100\times 300$. In both cases the rows of $X$ were generated as iid random vectors from the multivariate normal distribution $N(0,\Sigma)$, with $\Sigma=I$ in the left panel and $\Sigma$ in the second panel having a compound symmetry form with $\Sigma_{ii}=1$ and $\Sigma_{ij}=0.9$ for $i\neq j$. The x-axis represents the sparsity $k$ and the y-axis the fraction of sign vectors that satisfy the identifiability condition
(respectively, irrepresentability condition). In the left panel, the vertical lines represent the values $k = \frac{100}{2\log(300)}$ and $k=100\times \rho(100/300)$. These values correspond to the asymptotic upper bounds for $k$ so the irrepresentability and identifiability conditions hold for Gaussian design matrices with independent entries.}
\label{curve1}
\end{figure}

By definition, the  irrepresentability condition depends on $\beta$ by $S(\beta)$. Moreover, as asserted in Proposition \ref{sign}, the identifiability condition similarly depends only on $S(\beta)$. Thus, the comparison of these two conditions can be performed by considering parameter vectors such that $\beta=S(\beta)$. 
In Figure \ref{curve1}, we plot the irrepresentability and identifiability curves representing the proportion of sign vectors for which the identifiability condition or the irrepresentability condition is satisfied among $k-$sparse sign vectors. For each value of $k$, this proportion was estimated by generating 1000 sign vectors from a uniform distribution on the set $\{u\in \{-1,0,1\}^p\mid {\rm card}({\rm supp}(u))=k\}$.

Figure \ref{curve1} shows a large potential for improvement in sign detection by thresholded LASSO  compared to the base LASSO. If the columns of the design matrix are independent Gaussian vectors, then the irrepresentability condition is satisfied with high probability only if $k\leq 10$, which is also a limiting sparsity for the recovery of $S(\beta)$ by LASSO. Instead, the identifiability condition, which provides the limiting sparsity for the recovery of $S(\beta)$ by  the thresholded LASSO, holds as long as $k\le 30$. The simulations are in strong agreement with the asymptotic results in \cite{wainwright2009} and \cite{donoho_tanner}, which predict the upper bounds for $k$ such that the irrepresentability and identifiability conditions hold for Gaussian design matrices with independent entries. 

The difference between identifiability and irrepresentability curves becomes even more apparent when there are strong correlations between different columns in the design matrix. As expected, the irrepresentability curve shrinks towards zero. Instead, and rather unexpectedly, the simulated identifiability curve remains intact.

\subsection{Organization of the article}
 In Section 2, Theorem \ref{convergence} shows that identifiability is a necessary and sufficient condition for LASSO to separate the non-zero components of $\beta$ from the noise and to recover asymptotically 
$S(\beta)$ with thresholded LASSO and thresholded BPDN. 
Corollary \ref{cor:asymptotic} shows that under the asymptotic linear sparsity regime for Gaussian matrices thresholded LASSO and thresholded BPDN can recover the sign of $\beta$ when asymptotically 
 $k<n\rho(n/p)$. 
 
 In  Section 3, Proposition \ref{sign} shows that the identifiability condition depends only on 
$S(\beta)$ and not on the size of the non-zero components of $\beta$. Here we also introduce the irrepresentability and identifiability curves, which respectively give the fraction of $k-$sparse sign vectors satisfying the irrepresentability condition and the identifiability condition.
Section 4 is devoted to numerical experiments showing that sign estimators derived from the thresholded LASSO and the thresholded BPDN are better than sign estimators derived from LASSO and the adaptive LASSO, and that the knockoff method allows appropriate threshold selection for both methods.
The proofs are in the Appendix. In this section, we also formulate  Proposition 
\ref{CI} providing a tight upper bound on the probability of recovering $S(\beta)$ through LASSO. 

\subsection{Notation and assumptions}
\label{parametre}
In this article we always assume that  the design matrix $X$ has columns in  general position\footnote{Actually, general position is just a sufficient condition under which, for a fixed $X\in \R^{n\times p}$, independently of $Y\in \R^n$ the LASSO minimizer is unique. Recently, this condition was relaxed by Ewald and Schneider \cite{ewald2020distribution} to a geometric criterion which is both sufficient and necessary.} (see,  {\it e.g.}, \cite{tibshiraniryan} or the supplementary material for this manuscript). This assumption guarantees that the minimizer of (\ref{LASSOpen}) 
(resp. minimizer of (\ref{BPDN})) is unique and that therefore the LASSO estimator (resp. BPDN estimator) is well-defined. This assumption is very weak and generically holds. Indeed, when $X$ is a random matrix 
such that the entries $(X_{11},X_{12},\dots,X_{np})$ have a density on $\R^{np}$ then,
almost surely, $X$ is in general position \cite{tibshiraniryan}.\\

 Notation used in the following sections is as follows: 
\begin{itemize}
\item Let $I$ be a subset of $\{1,\dots,p\}$. We denote by $\overline{I}$ the complement of $I$, namely $\overline{I}:=\{1,\dots,p\}\setminus I$. 
\item The notation $X_I$ stands for a matrix whose columns are indexed by the elements of $I$: $(X_i)_{i\in I}$. 
\item For $b\in \R^p$, $b_I$ denotes the subvector containing elements of $b$ with indices in $I$. 
\item The symbols ${\rm supp}(b), {\rm supp}^+(b)$ and ${\rm supp}^-(b)$ denote, respectively, the sets 
$\{i \in \{1,\dots,p\} \mid b_i \neq 0\}, \{i \in \{1,\dots,p\} \mid b_i >0\}$ and 
$\{i \in \{1,\dots,p\} \mid b_i <0\}$.
\item LASSO and BPDN estimators depend on $X,\beta,\varepsilon$ and the tuning parameter $\lambda> 0$ or the regularization parameter $R\ge 0$. When appropriate, we use the parentheses to indicate these dependencies. The estimator $\widehat{\beta}$ indiscriminately represents the LASSO  estimator or the BPDN   estimator.
\end{itemize}

To formulate our asymptotic results, we will often consider a sequence of regression parameters $\beta^{(r)}\in \R^p$, $r\in \N$, 
for which non-zero components tend to infinity in the following way.

 \begin{assm}\label{assm1}
$\,$
\begin{itemize}
\item[{\bf 1)}] The sign of $\beta^{(r)}$ is invariant, namely there exists a sign vector $s^0 \in \{-1,0,1\}^p$ such that for any $r\in \N, S(\beta^{(r)})=s^0$.
\item[{\bf 2)}] The following limit holds $\lim_{r\to +\infty}\min\{|\beta^{(r)}_i|, i \in {\rm supp}(s^0)\}=+\infty$.
\item[{\bf 3)}] There exists $q> 0$ such that  
$$\forall r\in \N, \frac{\min\{|\beta^{(r)}_i|, i \in {\rm supp}(s^0)\}}{\|\beta^{(r)}\|_\infty}\ge q.$$ 
\end{itemize}
\end{assm}

\section{Identifiability is a necessary and sufficient condition for the sign recovery}
 
If $\beta$ does not satisfy the irrepresentability condition, then the LASSO sign estimator $S(\widehat{\beta}^{\rm L}(\lambda))$ is very unlikely to recover $S(\beta)$. However, one may relax this condition.
In fact,  Theorem \ref{convergence} shows that an appropriately thresholded version of LASSO (or thresholded version of BPDN)  recovers $S(\beta)$ if only the non-zero elements of $\beta$ are sufficiently large and the identifiability condition is satisfied.

\begin{thm}
\label{convergence}

Let $X$ be a $n\times p$ matrix in general position and such that ${\rm rank}(X)=n$ and let $\widehat{\beta}$ be the LASSO  or BPDN estimator with any fixed value of the tuning parameter $\lambda> 0$ or with any fixed regularization parameter $R\ge 0$. \\
{\bf Necessary condition for sign recovery:} If $S(\beta)$ is unidentifiable with respect to the $\ell_1$ norm, then the sign estimator derived from the thresholded LASSO or thresholded BPDN cannot recover $S(\beta)$. Indeed, for each fixed $\varepsilon 
\in \R^n$, 
 the sign of at least one non-zero component of $\beta$ is not correctly estimated 
 by $\widehat{\beta}(\varepsilon)$:
$$ \exists i \in {\rm supp}(\beta) \text{ \rm such that } \widehat{\beta}_i(\varepsilon)\beta_i\le 0. $$ 
{\bf Sufficient condition for sign recovery:} 
%This condition is asymptotic. 
Let $\beta^{(r)}$ be a sequence in $\R^p$ satisfying Assumption \ref{assm1}. 
If $s^0$ is identifiable with respect to the $\ell_1$ norm, then for any fixed $\varepsilon \in \R^n$ and sufficiently large $r> r_0(\varepsilon)$ the estimator $\widehat{\beta}(\varepsilon,r)$ separates negative components of $\beta^{(r)}$ ({\it i.e} $i\in {\rm supp}^-(\beta^{(r)})$), zero components of $\beta^{(r)}$ ({\it i.e} $i\notin {\rm supp}(\beta^{(r)})$) and positive components of $\beta^{(r)}$ 
 ({\it i.e} $i\in {\rm supp}^+(\beta^{(r)})$): 
\begin{itemize} 
\item[i)] $$ \forall i \in {\rm supp}(\beta^{(r)}),\,\, \widehat{\beta}_i(\varepsilon,r)\beta^{(r)}_i> 0. $$ 
\item[ii)]
$$\max_{i\in {\rm supp}^-(\beta^{(r)})}\left\{\widehat{\beta}_i(\varepsilon,r)\right\}<\min_{i\notin {\rm supp}(\beta^{(r)})}\left\{\widehat{\beta}_i(\varepsilon,r)\right\}\le \max_{i\notin {\rm supp}(\beta^{(r)})}\left\{\widehat{\beta}_i(\varepsilon,r)\right\}< \min_{i\in {\rm supp}^+(\beta^{(r)})}\left\{\widehat{\beta}_i(\varepsilon,r)\right\}.$$

\end{itemize}
\end{thm}

Let us note that the assumptions about $X$ are very weak and hold in general if $n \le p$.
The assumption that ${\rm rank}(X)=n$ ensures that the BPDN estimator is well-defined for any $R\ge 0$. The general position condition ensures uniqueness of both the LASSO and the BPDN estimator (see, {\it e.g.}, Proposition 1 in the Supplementary Material). 

Theorem \ref{convergence} emphasises that one cannot recover $S(\beta)$ with a sign estimator derived from LASSO or BPDN
 if $\beta$ is not identifiable with respect to the $\ell_1$ norm. If $\beta$ is identifiable with respect to the $\ell_1$norm, 
Theorem \ref{convergence} implies that $S(\beta)$ can be recovered by deriving sign estimators from the thresholded LASSO or thresholded BPDN.
In Section 4, we show how the appropriate thresholds can be obtained using control variables constructed by the knockoff method (see, {\it e.g.}, \cite{knockoffs, candes2016panning}).

In the asymptotic linear sparsity regime for Gaussian matrices, the transition curve $\rho(\cdot)$ described in \cite{DonTan05} 
allows us to characterize the identifiability condition, and so Theorem \ref{convergence} yields the Corollary \ref{cor:asymptotic}.
\begin{corr}
\label{cor:asymptotic}
Let $X$ be a $n\times p_n$ standard Gaussian matrix and let $k_n$ denote the number of non-zero components of $\beta^{(n)}\in \R^{p_n}$. \\
{\bf Necessary condition for sign recovery:}
If $n/p_n\rightarrow \delta \in(0,1)$, $k_n/n\rightarrow \xi \in(0,1)$ and $\xi>\rho(\delta)$ then asymptotically the sign of at least one non-zero component of $\beta^{(n)}$ is  incorrectly estimated by $\widehat{\beta}$:
$$\lim_{n\to +\infty}\P_{X,\varepsilon}\left(\exists i \in {\rm supp}(\beta^{(n)}) \text{ \rm such that } \widehat{\beta}_i\beta^{(n)}_i\le 0 \right)=1.$$
{\bf Sufficient condition for sign recovery:} 
%This condition is asymptotic on the signal strength. 
Given $n$, let $(\beta^{(n,r)})_{r\in \N}$ be a sequence satisfying the  Assumption \ref{assm1} (where $q$ does not depend on $n$).
 If $n/p_n\rightarrow \delta \in(0,1)$, $k_n/n\rightarrow \xi \in(0,1) $ and $\xi<\rho(\delta)$ then the estimator $\widehat{\beta}$ asymptotically separates the negative components of $\beta^{(n,r)}$, zero components of $\beta^{(n,r)}$ and positive components of $\beta^{(n,r)}$:
\begin{itemize} 
\item[i)] $$\lim_{n\to +\infty}\lim_{r\to +\infty}\P_{X,\varepsilon}\left(\forall i \in {\rm supp}(\beta^{(n,r)}),\,\, \widehat{\beta}_i\beta^{(n,r)}_i> 0 \right)=1.$$
\item[ii)]
$$\lim_{n\to +\infty}\lim_{r\to +\infty}\P_{X,\varepsilon}\left(\max_{i\in {\rm supp}^-(\beta^{(n,r)})}\left\{\widehat{\beta}_i\right\}<\min_{i\notin {\rm supp}(\beta^{(n,r)})}\left\{\widehat{\beta}_i\right\}\le \max_{i\notin {\rm supp}(\beta^{(n,r)})}\left\{\widehat{\beta}_i\right\}< 
\min_{i\in {\rm supp}^+(\beta^{(n,r)})}\left\{\widehat{\beta}_i\right\}\right)=1.$$

\end{itemize}
\end{corr}

\section{Identifiability and irrepresentability curves}
Given a certain design matrix $X$, we now define 
the irrepresentability indicator of the sign vector $s$.\\
{\bf Sign irrepresentability indicator:}
$$\Phi^X_{\rm IC }: s\in \{-1,0,1\}^p \mapsto \begin{cases} 1 \text{ \rm if } s=(0,\dots,0)\\
1 \text{ \rm if } \ker(X_I)=\{{\bf 0}\} \text{ \rm and } \|X_{\overline{I}}'X_I(X_I'X_I)^{-1}s_I\|_\infty \le 1 \text{ \rm where }
I:={\rm supp}(s) \\ 
0 \text{ \rm otherwise } \end{cases}.$$
This irrepresentability indicator shows whether the LASSO sign estimator can recover $S(\beta)$. Namely, if $\phi^X_{\rm IC }(S(\beta))=0$ then $S(\beta)$ cannot be recovered using the LASSO sign estimator, even if the non-zero components of $\beta$ are extremely large. 

 Proposition \ref{sign} shows that the identifiability condition also depends only on $S(\beta)$ and not on the sizes of the non-zero components of $\beta$.

\begin{prop} \label{sign}

Consider two vectors $b\in \R^p$ and $\tilde{b} \in \R^p$ such that $S(b)=S(\tilde{b})$. Then $\tilde{b}$ is identifiable with respect to the matrix $X$ and the $\ell_1$ norm if and only if $b$ is identifiable with respect to the matrix $X$ and the $\ell_1$ norm.

\end{prop}
Given a certain design matrix $X$, the sign identifiability indicator is defined as follows. \\
{\bf Sign identifiability indicator:} 
$$\Phi^X_{\rm Idtf}: s \in \{-1,0,1\}^p \mapsto \begin{cases} 0 \text{ \rm if } 
s \neq \underset{b \in \R^p}{\text{ \rm argmin }} \|b\|_1 \text{ \rm subject to }Xb=Xs  \\
1 \text{ \rm otherwise } \end{cases}.$$
This identifiability indicator shows whether the sign estimators obtained by thresholded LASSO and thresholded BPDN can recover $S(\beta)$. Namely, if $\phi^X_{\rm Idtf}(S(\beta))=0$ then the thresholded LASSO (respectively, thresholded BPDN) 
sign estimator fails to recover $S(\beta)$ even if the non-zero components of $\beta$ are extremely large.

According to Proposition 2 in the Supplementary Material, $\beta$ does not satisfy the identifiability condition if the columns $(X_i)_{i\in {\rm supp}(\beta)}$ are not linearly independent.
Consequently, if ${\rm card}({\rm supp}(\beta))> n$ then 
$\phi^X_{\rm IC }(S(\beta))=\phi^X_{\rm Idtf}(S(\beta))=0$. Let us give some basic properties and comments about the two indicator functions. 
\begin{enumerate}
\item Both $\phi^X_{\rm IC }$ and $\phi^X_{\rm Idtf}$ are even.
\item Given Proposition \ref{CI_implique_NSP}, for each $s\in \{-1,0,1\}^p$, 
$\Phi^X_{\rm IC }(s)\le \Phi^X_{\rm Idtf}(s)$. 
\item The computation of $\Phi^X_{\rm IC }$ requires only a simple matrix calculus;
the computation of $\Phi^X_{\rm Idtf}$ requires only solving the basis pursuit problem. 
\end{enumerate}

\subsection{Graphs of identifiability and irrepresentability curves}
\label{subsection}

\begin{figure}[htbp]
\centering
\includegraphics[scale=0.75]{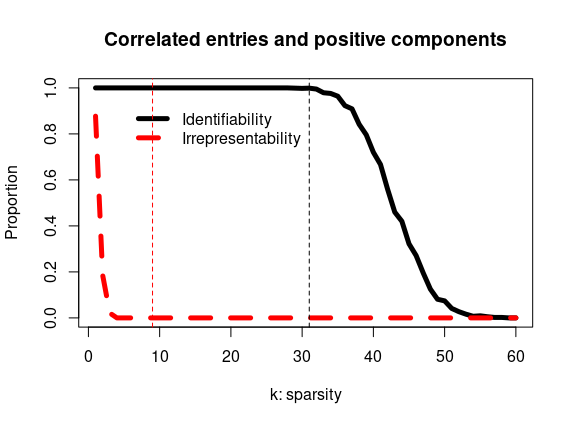}

	\caption{Graphs of functions $k \mapsto p^X_{\rm Idtf+}(k)$ and $k \mapsto p^X_{\rm IC +}(k)$ where $X$ is described in the setting above. The vertical lines represent the values $k = \frac{100}{2\log(300)}$ and $k=100\times \rho(100/300)$.}
	\label{Idtf_vs_IC_positif}
\end{figure}

The number of sign vectors is very large ($3^p$) and therefore we cannot explicitly specify $\Phi^X_{\rm Idtf}$ and $\Phi^X_{ IC }$ for each sign vector. Instead, we define the identifiability and irrepresentability curves as the following functions of the sparsity $k$ of the vector $\beta$, $k={\rm card}({\rm supp}(\beta)) \in \{1,\dots,n\}$:
\begin{itemize}
\item identifiability curve is defined as $p^X_{\rm Idtf}(k):=\E_U(\Phi^X_{\rm Idtf}(U))$,
\item irrepresentability curve is defined as $p^X_{\rm IC }(k):=\E_U(\Phi^X_{ IC }(U))$,
\end{itemize}
where $U$ is uniformly distributed on $\{u\in \{-1,0,1\}^p \mid {\rm card}({\rm supp}(u))=k\}$.
Figure \ref{curve1} in the Introduction provides identifiability curves $k\in \{1,\dots,60\}\mapsto p^X_{\rm Idtf}(k)$ and irrepresentability curves $k\in \{1,\dots,60\}\mapsto p^X_{\rm IC }(k)$ for two specific $100\times 300$ matrices (one generated with iid $\mathcal{N}(0,1)$ entries and the other generated with positively correlated $\mathcal{N}(0,1)$ entries).
In addition, for the case where the design matrix $X$ has positively correlated entries, we also consider a situation where $U$ is uniformly distributed on $\{u\ \in \{0,1\}^p \mid {\rm card}({\rm supp}(u))=k\}$. Specifically, we consider the following setting:

\begin{description}
\item[Positively correlated entries and positive components:]  $X$ is a fixed design matrix with $n=100$, $p=300$, whose rows are generated by independent draws from the multivariate normal distribution 
$\mathcal{N}({\bf 0},\Gamma)$, with 
$\Gamma_{ii}=1$ for $i \in \{1,\dots,p\}$ and $ \Gamma_{ij}=0.9$ when $i \neq j$ (the same matrix as the one used for the right panel in Figure \ref{curve1}). The distribution of the sign vectors is uniform on 
$\{u\in \{0,1\}^p \mid {\rm card}({\rm supp}(u))=k\}$. 
\end{description}
\noindent
Figure \ref{Idtf_vs_IC_positif} shows an interesting behavior of the irrepresentability and identifiability curves ($k \mapsto p^X_{\rm IC +}(k)$ and $k \mapsto p^X_{\rm Idtf+}(k)$) in the above setting. Here we  observe that the irrepresentability condition becomes even more stringent than in the case when the distribution of the elements of the sign vector is symmetric. At the same time, the identifiability condition now becomes much weaker and is satisfied under a much wider range of sparsity levels  compared to the identifiability curve in the right panel of Figure \ref{curve1}. 

\section{Numerical comparisons of sign estimators} 
Theorem \ref{convergence} states that the sign estimators provided by thresholded LASSO or thresholded BPDN can recover ${\rm S}(\beta)$ as long as the identifiability condition is satisfied. Another way to recover ${\rm S}(\beta)$ is to use a sign estimator provided by the adaptive LASSO proposed in \cite{zou}. 
Indeed, as claimed in \cite{zou} or \cite{huang}, if the weights for adaptive 
LASSO are based on an accurate estimator for $\beta$ one  obtains a sign estimator that is consistent for ${\rm S}(\beta)$ under much weaker assumptions than the irrepresentability condition.
In this section, we numerically compare the sign estimators obtained by LASSO, thresholded LASSO, thresholded BP and adaptive LASSO. We also include a comparison with LASSO-zero  \cite{descloux}, which is a modification of the thresholded BP aimed at controlling the number of false discoveries.

\subsection{Selection of the tuning parameter}

As explained in \cite{AMP_knockoffs,wang}, a value of the optimal tuning parameter for sign recovery by thresholded LASSO is much smaller than the optimal value of the tuning parameter for LASSO sign estimator. Specifically:
\begin{itemize}
\item For the LASSO sign estimator, the tuning parameter must be large enough to prevent the inclusion of false detections.

\item For the thresholded LASSO sign estimator, the tuning parameter should be chosen to achieve good separation between zero and non-zero elements of $\beta$. This requires good estimation rather than good selection properties. Here, the tuning parameter does not need to be large, as the threshold allows false detections to be eliminated.

\end{itemize}

\subsubsection{The tuning parameter for the LASSO sign estimator}
If  $S(\beta)$ satisfies the irrepresentability condition, then Proposition \ref{CI} in the Appendix implies that one can choose a tuning parameter 
$\lambda_L$ such that, for sufficiently large $\beta$, $\P(S\left(\beta(\lambda_L)\right)=S(\beta))$ is arbitrarily close to a given value from  $(0,1)$ ({\it e.g.} $0.95$). 
According to the irrepresentability curve in Figure \ref{curve1}, the irrepresentability condition is satisfied with probability close to 1 if the elements of $X$ are iid $\mathcal{N}(0,1)$ random variables and $\beta$ contains $k=5$ non-zero elements.
Therefore, in this setting, we  choose $\lambda_L$ such that the probability of sign recovery converges to 0.95 and the probability of at least one false discovery (Family Wise Error Rate, FWER) converges to 0.05 when the signal magnitude tends to infinity.
Since the irrepresentability condition is  not satisfied for the remaining scenarios in our simulation study and thus the FWER cannot be controlled at a low level, the performances of LASSO are not reported.

%use the same value $\lambda_L=81.18$ for all our %simulations.
\subsubsection{The tuning parameter for the thresholded LASSO sign estimator}
The tuning parameter can be chosen using the asymptotic theory of the Approximate Message Passing (AMP) algorithm for LASSO provided, for example, in \cite{bayati2012, su2017false, AMP_knockoffs}.
In this theory, the elements of the design matrix are iid Gaussian $\mathcal{N}(0,1)$  variables and the components of $\beta$ are iid random variables from the mixture distribution $\Pi=(1-\gamma) \delta_{0}+\gamma \Pi^{\star}$, where $\delta_0$ is a point mass distribution concentrated at 0 and $\Pi^{\star}$ is an arbitrary fixed distribution. The asymptotic  mean squared error of LASSO is derived under the assumption that the number of observations $n$ and the number of explanatory variables $p$ tend to infinity and that 
$n/p$ tends to $ \delta > 0$.
Then  the tuning parameter $\lambda_{ AMP }$  is chosen such that the asymptotic mean squared error is minimal (see, {\it e.g.}, the prescription in \cite{AMP_knockoffs, wang}).
As discussed in \cite{AMP_knockoffs, wang}, such a tuning parameter for any fixed value of Type I error allows maximizing the asymptotic power of the  LASSO. In our simulation study, we used this asymptotically optimal $\lambda_{ AMP } (k,t)$ for the parameter values:
 $\delta=n/p=100/300$, $\gamma=k/p=k/300$ and $\Pi^{\star}=1/2\delta_t+1/2\delta_{-t}$,
where $\delta_{t}$ is a point mass distribution at $t$. The values $\lambda_{ AMP } (k,t)$ were then used for the simulations in the independent case.
In the ``correlated'' case, these values proved to be sub-optimal, so in this case we additionally obtained results for tuning parameters equal to $0.5 \lambda_{ AMP } (k,t)$, which provide much better empirical performance.

\subsection{Selection of the threshold}
We define the thresholded LASSO estimator (or the thresholded BP estimator) to be 
\begin{equation}\label{BP_thres}
\forall i\in \{1,\dots,p\}, \widehat{\beta}^\tau_i:=\widehat{\beta}_i{\bf 1}_{\left\{\left|\widehat{\beta}_i\right|>\tau\right\}}\;.
\end{equation}
Given a threshold $\tau> 0$, we now define FWER as$${\rm FWER}(\tau):=\P\left(\exists i \notin {\rm supp}(\beta),\left|\widehat{\beta}^{\tau}_i\right| \neq 0\right).$$

By taking $\tau_{\alpha}$ as the $1-\alpha$
quantile of the distribution of $\max\left\{|\widehat{\beta}_i|, i \notin {\rm supp}(\beta)\right\}$ we would control FWER exactly at the $\alpha$ level. 
However, $\tau_{\alpha}$ cannot be readily determined since $\beta$ is not known.

To obtain a threshold greater than $\tau_{\alpha}$ (and thus control for FWER at the $\alpha$ level), it seems appealing to consider the distribution of the supremum norm of the LASSO estimator (or BP estimator) in the complete null model when $\beta={\bf 0}$
{\cite{giacobino}}.
For the BP estimator, Descloux and Sardy \cite{descloux} propose the threshold $\tau^{\rm fn}_{\alpha}$ 
 defined as the $1-\alpha$ quantile of $\max\left\{\left|\widehat{\beta}_1^{\rm fn}\right|,\dots,\left|\widehat{\beta}_p^{\rm fn}
\right|\right\}$ where $\widehat{\beta}^{\rm fn}$ is the following estimator
$$\widehat{\beta}^{\rm fn}:=\text{ \rm argmin }\|\beta\|_1 \text{ \rm subject to }X\beta=\varepsilon,
\text{ \rm where $\varepsilon \sim \mathcal{N}_n(0,\sigma^2 I)$}.$$
Unfortunately, if the vector $\beta$ contains some non-zero elements, this intuitive method yields a threshold $\tau^{\rm fn}_{\alpha}$ smaller than $\tau_{\alpha}$ and thus ${\rm FWER}(\tau^{\rm fn}_{\alpha})>\alpha$
(see also Su et al. \cite{su2017false} for further explanation).

The recently developed knockoff methodology \cite{knockoffs, candes2016panning} provides control of False Discovery Rate (FDR).
	This control is achieved by adding additional control variables to the design matrix. 
	Originally designed to control the FDR, the control variables also allow us to approximate the distribution of estimators 
	corresponding to the zero components of $\beta$. 
In this numerical study, we informally use the model-free knockoffs proposed in \cite{candes2016panning} to approximate a threshold that controls the FWER at a certain level.
The approach developed below is suitable for the situation when $X$ is a random matrix whose distribution is invariant to the permutation of the columns. In this setting, we can generate the knockoff variables one by one instead of generating the  full $n\times p$ knockoff matrix (see Weinstein et al. \cite{weinstein2017power} for a similar approach). Since adding the controlled variables may change some relevant properties ({\it e.g.} the identifiability condition for $\beta$), we should ideally add only one knockoff variable at a time when computing LASSO estimates. 
However, this would lead to a heavy computational burden on the procedure for estimating the relevant threshold. Therefore, in our simulation study, we use model-free knockoffs \cite{candes2016panning, weinstein2017power} to generate $30=p/10$ controlled variables. Then Lasso or BP is run on the matrix augmented with these additional columns and the maximum of the absolute values of the regression coefficients over 30 controlled variables is stored. This step is repeated $10$ times and the total maximum of $p=300$ absolute values of regression coefficients over controlled variables is calculated. The whole procedure is repeated many times (1000 in this case) and the 0.95-quantile of the obtained maxima is used as a threshold to identify zero components of $\beta$.
To be consistent with the setup of the simulations used to derive the irrepresentability and identifiability curves, we used the same two fixed design matrices $X$ as in Figure \ref{curve1}, while the positions of the $k$ sparse signals and the error terms were randomly generated for each of the 1000 replicates.

\subsubsection{LASSO and Adaptive LASSO}
In our numerical experiments, we chose the following values of tuning parameters for LASSO and adaptive LASSO:
\begin{itemize}
\item For LASSO, we chose $\lambda_L=81.18$ to control for FWER at the 0.05 level when $k=5$ and the covariates are independent.
\item For adaptive LASSO, the weights are derived using initial estimates $\widehat{\beta}^L(\lambda_{ AMP })$, where the tuning parameter is chosen according to the AMP theory, described above. For $i\in \{1,\dots,p\}$, the weights $w_i$ are defined as 
$w_i:=1/(|\widehat{\beta_i}^L(\lambda_{ AMP })|+10^{-7})$.
With these weights and the tuning parameter $\lambda_L$ described above, the adaptive LASSO has the following expression
\begin{equation}
\label{LASSO_adapt}
\widehat{\beta}^{\rm adapt}:=\underset{\beta \in \R^p}{\text{\rm argmin }}\frac{1}{2}\|Y-X\beta\|_2^2+\lambda_L \sum_{i=1}^{p}
w_i|\beta_i|.
\end{equation}
\end{itemize}
In all our simulations, LASSO is computed using {\it glmnet} R package.

\subsubsection{ LASSO-zero from \cite{descloux}}
Given iid $n\times q$ standard Gaussian matrices $G^{(1)},\dots,G^{(M)}$, we consider the following basis pursuit minimizers:
$$\forall k\in \{1,\dots,M\}, (\hat \beta^{(k)}, \hat \gamma^{(k)}):=\underset{b\in \R^p, c \in \R^q}
{\text{\rm argmin }}\|b\|_1+\|c\|_1 \text{ \rm subject to }
y=Xb+G^{(k)}c.$$
%From $\hat \beta^{(1)},\dots,\hat \beta^{(M)}$ andmessages
Now, given a threshold $\tau> 0$, we can define the LASSO-zero estimator $\hat\beta^{{\rm lass0}(q,M)}_\tau$ as follows:
$$\forall i \in \{1,\dots,p\}, \hat{\beta}^{med}_i={\rm median}\{\hat\beta^{(1)}_i,\dots,\hat\beta^{(M)}_i\}
\text{ \rm and }\hat\beta_{\tau,i}^{{\rm lass0}(q,M)}=\hat\beta^{med}_i{\bf 1}(|\hat\beta^{med}_i|> \tau).$$
One can use the knowledge that $\sigma=1$, to compute the QUT threshold $\tau$ as the $1-\alpha$ quantile of $\|\hat \beta^{med}\|_\infty$ when $\beta={\bf 0}$. Otherwise, one can also pivotize the $\|\hat \beta^{med}\|_\infty$ statistic to compute this threshold (see \cite{descloux} for details).

\subsection{Numerical comparisons}

All numerical experiments are performed with the two $100\times 300$ design matrices $X$ already used in Figure \ref{curve1} and in Section \ref{subsection}.
We set $\beta\in \R^p$ such that $k:={\rm card}({\rm supp}(\beta))$ with $k\in\{5,20\}$ 
and the $k$ elements of ${\rm supp}(\beta)$ are drawn without replacement from $\{1,\dots,p\}$. The non-zero components of $\beta$ are sampled from the distribution $P(\beta_i=t)=P(\beta_i=-t)=0.5$ with $t> 0$. Additionally, when the columns of $X$ are correlated, we consider the constellation in which all non-zero coefficients are equal to $t$. In all simulations, the error term is generated as $\varepsilon\sim\mathcal{N}(0,Id_n)$.

\begin{figure}[htbp]
\begin{tabular}{c c}
\includegraphics[scale=0.55]{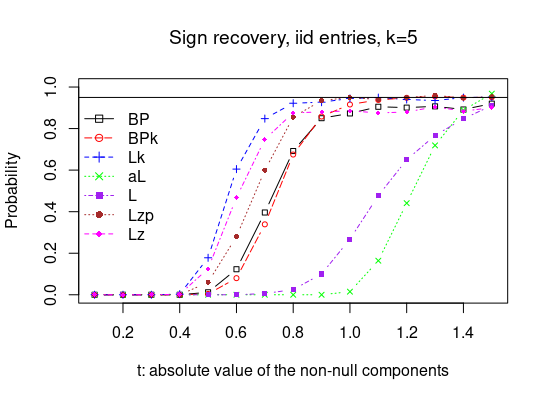}& 
\includegraphics[scale=0.55]{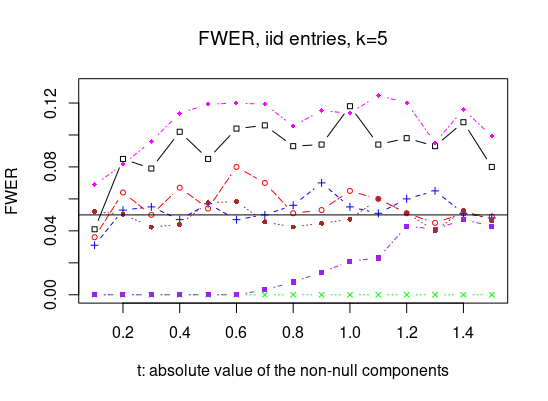}\\
\includegraphics[scale=0.55]{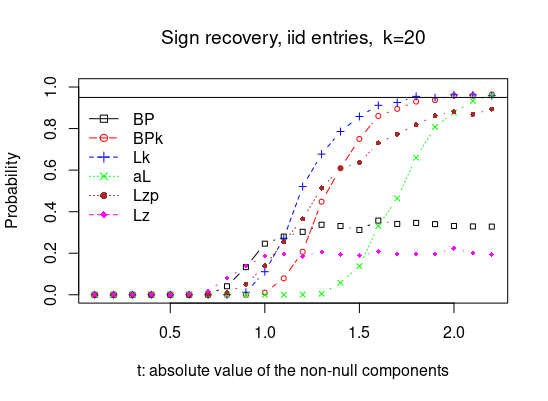}&
\includegraphics[scale=0.55]{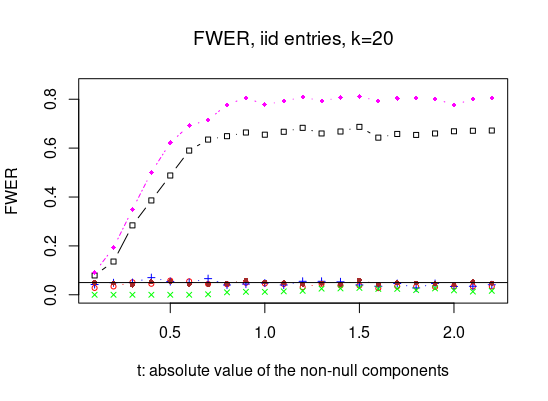}\\
\end{tabular}
\caption{ This figure shows the FWER and the probability of recovering $S(\beta)$ for each sign estimator and when $X$ contains iid $\mathcal{N}(0,1)$ entries. The graphs on the left show the probability of recovering $S(\beta)$ (on the y-axis) as a function of $t$ (on the x-axis), where $t$ indicates how large the non-zero components of $\beta$ are.
The graphs on the right show the FWER (on the y-axis) as a function of $t$ (on the x-axis). 
}
\label{ind}
\end{figure}

\begin{figure}[htbp]
\begin{tabular}{c c}
\includegraphics[scale=0.55]{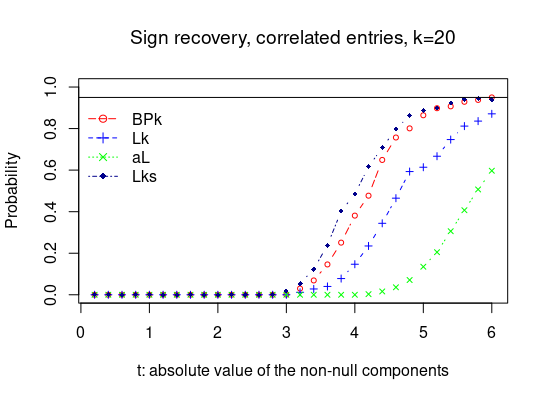}& 
\includegraphics[scale=0.55]{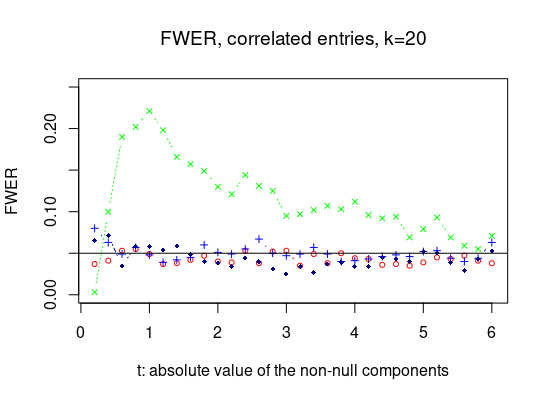}\\
\includegraphics[scale=0.55]{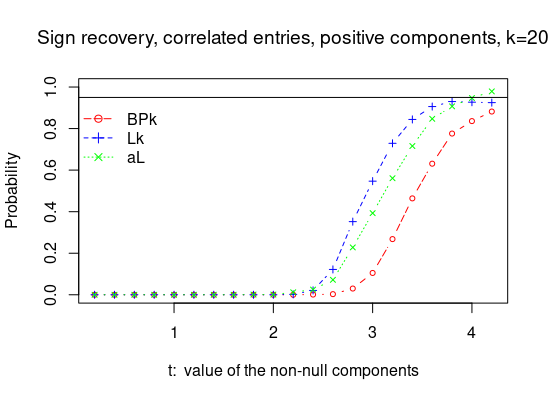}&
\includegraphics[scale=0.55]{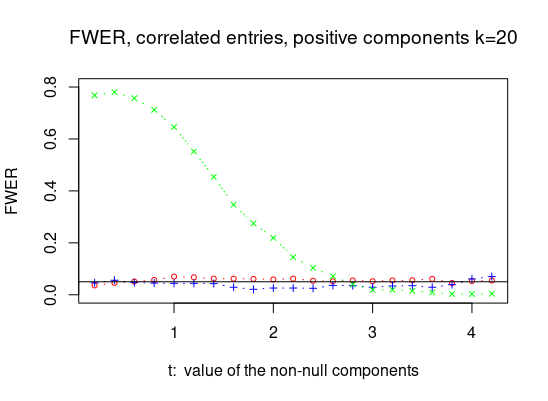}\\
\end{tabular}	
\caption{This figure shows the FWER and the probability of recovering $S(\beta)$ for each sign estimator when $X$ has strongly correlated columns. The plots on the left show the probability of recovery of 
$S(\beta)$ (on the y-axis) as a function of $t$ (on the x-axis), where $t$ measures how large the non-zero components of $\beta$ are. The graphs on the right show the FWER (on the y-axis) as a function of $t$.}
\label{corr_same}

\end{figure}

Figures \ref{ind}-\ref{corr_same} show the comparison between the following sign estimators.
\begin{itemize}
\item The sign estimator {\bf L} from LASSO with $\lambda=\lambda_L$.
\item The sign estimator {\bf aL} from the adaptive LASSO, described in (\ref{LASSO_adapt}).
\item The sign estimator {\bf BPk} from the thresholded BP, where the threshold is given by the "knockoff" method described above.
\item The sign estimator {\bf Lk} from the thresholded LASSO with $\lambda=\lambda_{ AMP }$ and with a threshold given by the ``knockoff'' method described above.
\item The sign estimator {\bf Lks} from the thresholded LASSO with $\lambda=0.5 \lambda_{ AMP }$ and with a threshold given by the ''knockoff'' method described above.
\item The sign estimator {\bf Lz} of LASSO-zero, where $M=30, q=n$. The QUT threshold is not data driven and is computed as the $0.95$ quantile of $\|\hat \beta^{med}\|_\infty$ when $\beta={\bf 0}$. 
\item The sign estimator {\bf Lzp} of LASSO-zero when $M=30, q=n$. The QUT threshold is data-driven and is computed by pivotizing the statistic $\|\hat \beta^{med}\|_\infty$, as explained in \cite{descloux}. 
\end{itemize}

We give the curves representing the following statistical properties as a function of $t> 0$:
\begin{itemize}
\item {\bf Probability} is the proportion of 1000 replicates for which the sign is recovered.
\item {\bf FWER} is the proportion of 1000 replicates for which at least one zero component of $\beta$ is not estimated to be zero.
\end{itemize}  
 FWER is directly related to the probability of sign recovery, namely $FWER\geq \alpha $ implies that the probability of sign recovery is less than $1-\alpha$. 
The sign estimators {\bf L}, {\bf BPk}, {\bf Lk} and {\bf Lks} were specifically designed to control for FWER at the 0.05 level.

If $k=5$ and the elements of $X$ are iid normal variables, then the irrepresentability condition holds and LASSO can recover the true model. Figure \ref{ind} shows that in this case the upper bound on the probability of recovering the sign given in Proposition \ref{CI} is achieved by LASSO and the FWER is controlled. However, for moderately large signals, the probability of recovering $S(\beta)$ by LASSO is much smaller than by the thresholded versions of BP and Lasso, as well as by Lasso-zero.
If $k=20$,  the irrepresentability condition is not satisfied, but the identifiability condition holds, thus the thresholded LASSO and the thresholded BP  identify the sign of $\beta$ when non-zero elements are large and the threshold is properly calibrated.

Figure \ref{ind} shows that thresholds for {\bf BP } or {\bf Lz}, which are not data-driven and are computed in the complete null model ({\it i.e.} when $\beta={\bf 0}$) do not control the FWER when $\beta$ has many large components (intuitively, when $\beta$ is far from ${\bf 0}$). Consequently, {\bf BP } or {\bf Lz} cannot recover $S(\beta)$ with a large probability.
Instead, our heuristic application of the knockoff method, as well as LASSO-zero when the threshold is data-driven, allow us to almost perfectly control FWER at $0.05$. Consequently, when the non-zero components of $\beta$ are large enough, the sign estimators derived from {\bf Lk}, {\bf BPk}, and {\bf Lzp} recover $S(\beta)$ with probability close to $0.95$. Among these methods, the thresholded LASSO estimator {\bf Lk} with $\lambda$ selected by AMP theory systematically has the highest probability of recovering $S(\beta)$ for moderately large signals.

Figure \ref{corr_same} shows the results of the simulation study for the design matrix $X$ with very highly correlated columns. In this setting, the probability of recovering $S(\beta)$ by LASSO-zero was very close to zero, and for clarity of presentation, LASSO-zero is not included in Figure \ref{corr_same}.

Figure \ref{corr_same} confirms the numerical experiments given in Figure \ref{ind}. In particular, 
thresholded BP and thresholded LASSO 
 can recover
$S(\beta)$ when the threshold is well calibrated and the non-zero components of $\beta$ are large enough. If the tuning parameter is chosen correctly, thresholded LASSO recovers $S(\beta)$ with a higher probability than thresholded BP. Similar to the case of independent entries, our heuristic knockoff approach allows to  control FWER at the 0.05 level.

\subsection{Numerical experiments with the design matrix from the riboflavin dataset}
\begin{figure}[htbp]
\centering
\includegraphics[scale=0.75]{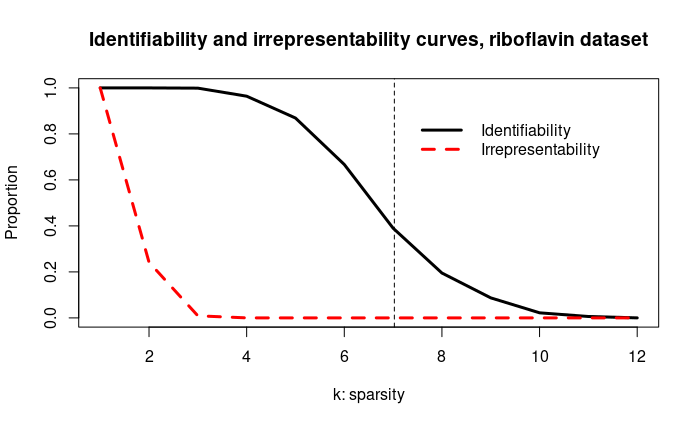}
	\caption{Graphs of the functions $k \mapsto p^X_{\rm Idtf}(k)$ and  $k \mapsto p^X_{\rm IC}(k)$ where $k\in \{1,\dots,12\}$ and  $X$ is the matrix from the riboflavin dataset. The equation of the vertical line is $k=n\rho(n/p)$.}
	\label{Idtf_vs_IC_ribo}
\end{figure}

In this section, we consider the same simulation setting as in \cite{descloux}, which deals with the riboflavin dataset \cite{buhlmann2014high}.
The design matrix $X$ contains expression levels of $p = 4088$ genes for $n = 71$ Bacillus subtilis bacteria. The matrix is mean-centered (for each $j\in \{1,\dots,p\}, \sum_{i=1}^{n}X_{ij}=0$) and standardized (for each $j\in \{1,\dots,p\}, \frac{1}{n-1}\sum_{i=1}^{n}X_{ij}^2=1$). Figure \ref{Idtf_vs_IC_ribo}
shows irrepresentability and identifiability curves of this design matrix.

 The identifiability and irrepresentability curves shown in Figure \ref{Idtf_vs_IC_ribo} confirm the numerical experiments reported in Descloux and Sardy \cite{descloux}. In particular, the probability of sign recovery by LASSO reported in panel (c) of Figure 6 in \cite{descloux} is below the irrepresentability curve from Figure \ref{Idtf_vs_IC_ribo}
 and the probability of sign recovery by LASSO-zero 
 is below the identifiability curve.
 
 We also ran the simulations comparing LASSO-zero and thresholded LASSO on the riboflavin dataset. In this study, we set $k:={\rm card}({\rm supp}(\beta))\in \{1,2,3\}$ (so that the identifiability condition holds),
and the non-zero coefficients are set to 2 with random signs. In addition to the probability of sign recovery and the FWER, we also estimate 
 the False Discovery Rate (FDR) and the True Positive Rate (TPR).

\begin {itemize}
\item {\bf FDR} is the mean of 1000 replicates of the False Discovery Proportion (FDP), where the FDP is defined as follows:
$$FDP=\frac{{\rm card}(\{i \notin {\rm supp}(\beta)| \hat\beta_i\neq 0\})}{\max\{{\rm card}({\rm supp} (\hat\beta)),1\}}.$$
\item {\bf TPR} is the mean of 1000 replicates of the True Positive Proportion (TPP), where the TPP is defined as follows:
$$TPP=\frac{{\rm card}( {\rm supp}(\beta)\cap {\rm supp}(\hat\beta))}{{\rm card}({\rm supp}(\beta))}.$$

\end{itemize} 
For the thresholded LASSO, we selected the tuning parameter by cross-validation. To estimate the threshold, we used the heuristic knockoff procedure described above, where for each run of LASSO we added $q=p/8=511$ columns to the design matrix $X$, drawn from the multivariate normal distribution $N(0,\Gamma)$, where $\Gamma_{ii}=1$ and $\Gamma_{ij}=\frac{1}{p(p-1)(n-1)} \sum_{i\neq j} X_i' X_j=0.0292$.

\begin{figure}[htbp]
\begin{tabular}{c c}
\includegraphics[scale=0.55]{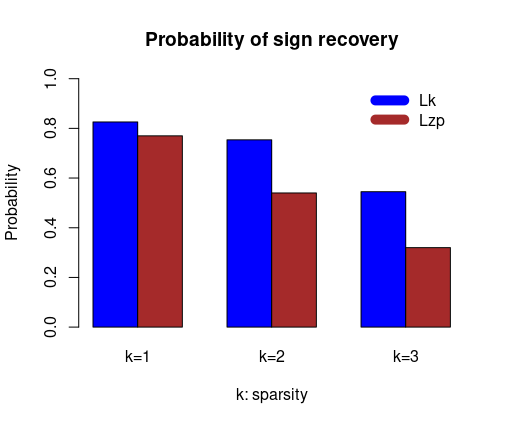}& 
\includegraphics[scale=0.55]{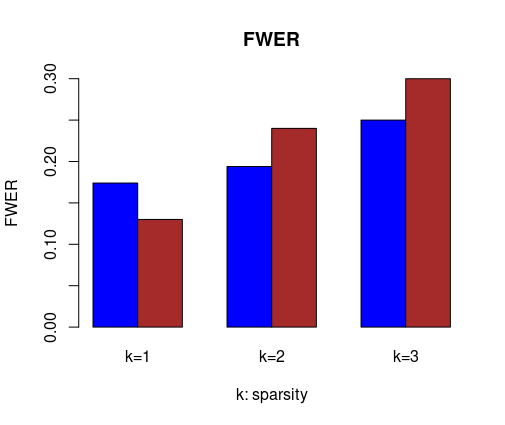}\\
\includegraphics[scale=0.55]{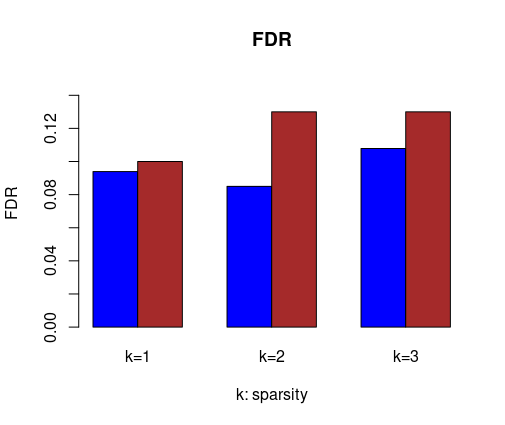}&
\includegraphics[scale=0.55]{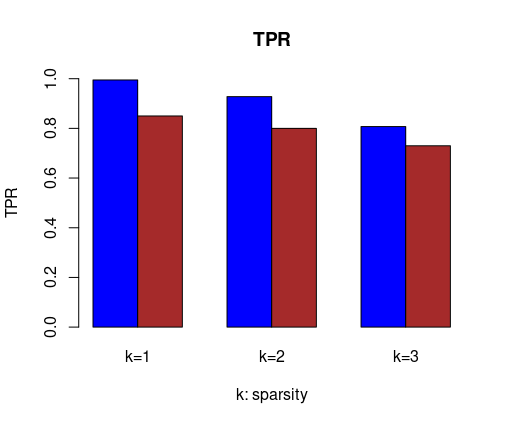}\\
\end{tabular}	
\caption{This figure provides the probability of sign recovery, FWER, FDR and TPR of LASSO-zero ({\bf Lzp})  and thresholded LASSO ({\bf Lk}) as a function of the sparsity $k\in \{1,2,3\}$.}
\label{riboflavin}
	
\end{figure}
Figure \ref{riboflavin} shows that LASSO-zero and thresholded LASSO have FWER above the assumed value of 0.05. This is not surprising since the gene expressions are not exactly normally distributed. However, the FWER of both methods is still quite low (below 0.3) and the FDR is kept close to 0.1. 
Since the correlations between columns in the design matrix are not too large, the thresholded LASSO and LASSO-zero show similar performance.
However, the probability of sign recovery and TPR is systematically larger for thresholded LASSO, which also has smaller FWER and FDR.

\section{Conclusion}
The focus of this article is on the theoretical properties of sign estimators derived from LASSO, thresholded LASSO, and thresholded BPDN . 
When $S(\beta)$ is identifiable with respect to the $\ell_1$ norm and when non-zero components of $\beta$ tends to infinity, we have shown that sign estimators derived from thresholded LASSO and thresholded BPDN recover $S(\beta)$. On the other hand, if $S(\beta)$ is not identifiable with respect to the $\ell_1$ norm, sign estimators derived from thresholded LASSO and thresholded BPDN cannot recover $S(\beta)$.

We introduced irrepresentability and identifiability curves that provide information about the probability of sign recovery by LASSO or thresholded LASSO and thresholded BPDN as a function of the number of non-zero elements in the vector of regression coefficients.
 
The performances of the sign estimators derived from LASSO, thresholded LASSO and thresholded BPDN obviously depend on the tuning parameters and the threshold. We have illustrated that AMP theory and the knockoff method are useful to select these parameters. Our simulations show that thresholded LASSO and thresholded BPDN sign estimators  outperform the adaptive LASSO and LASSO sign estimators.

\section*{Acknowledgments}

We would like to thank Emmanuel J. Cand\`es and Wojciech Rejchel for helpful comments. Ma\l{}gorzata Bogdan's research was supported by NCN grant 2016/23/B/ST1/00454. We gratefully acknowledge the grant from the Wroc\l{}aw Center of Networking and Supercomputing (WCSS), where most of the computations were performed.

\section{Appendix}

\subsection{Sign recovery with LASSO sign estimator}
The upper bound for sign recovery given in Proposition \ref{CI} appears in Lemma 3 of Wainwright 
\cite{wainwright2009} as a technical result establishing the irrepresentability condition (Theorem 2 in Wainwright 
\cite{wainwright2009}). According to Proposition \ref{CI}, if $\beta$ is identifiable with respect to the $\ell_1$-norm, this upper bound is asymptotically reached as soon as $\min \{ |\beta_i|, i\in {\rm supp}(\beta)\}$ tends to $+\infty$.

\begin{lemm}[Lemma 3 from \cite{wainwright2009} ]\label{Wein}
Let $I:={\rm supp}(\beta)$ and let 
$X_I, X_{\overline{I}}$ be matrices whose columns are $(X_i)_{i\in I}$ and $(X_i)_{i\notin I}$, respectively. 
Let us assume that $\ker(X_I)=\{{\bf 0}\}$ and let  $\zeta_{X,\lambda,S(\beta)}:=X_{\overline{I}}'X_I(X_I'X_{I})^{-1}S(\beta_I)+ \frac{1}{\lambda}
X_{\overline{I}}'\left(Id-X_I(X_I'X_I)^{-1}X_I'\right)\varepsilon.$ 

The following upper bound for the sign recovery holds.
$$\P\left(S(\widehat{\beta}^{\rm L}(\lambda))=S(\beta)\right)\le 
\P\left(\left\|\zeta_{X,\lambda,S(\beta)}\right\|_\infty \le 1\right).$$ 
\end{lemm}

\begin{prop}
\label{CI}
Assume that the assumptions of Lemma \ref{Wein} hold. Moreover,
assume that the sequence $(\beta^{(r)})$ in $\R^p$  satisfies  Assumption \ref{assm1}.
If $s^0$ is identifiable with respect to the $\ell_1$-norm, then the following asymptotic results hold.
\begin{description}
\item[Sharpness of the upper bound:] Asymptotically, the upper bound is obtained.

\begin{eqnarray*}
&&\limsup_{r\to+\infty}\P\left(S(\widehat{\beta}^{\rm L}(\lambda,r))=s^0\right)\le  \P\left(\left\|\zeta_{X,\lambda,s^0}\right\|_\infty \le 1\right),\\
&& \liminf_{r\to+\infty}\P\left(S(\widehat{\beta}^{\rm L}(\lambda,r))=s^0\right)\ge  
\P\left(\left\|\zeta_{X,\lambda,s^0}\right\|_\infty < 1\right).
\end{eqnarray*}
\color{black}
\item[Asymptotic control of FWER:]Let us define $\P\left(\left\|\zeta_{X,\lambda,s^0}\right\|_\infty <1\right)=\gamma$ and 
$\P\left(\left\|\zeta_{X,\lambda,s^0}\right\|_\infty \le 1\right)=\bar{\gamma}$.
The sign of the non-zero elements of $\beta^{(r)}$ is correctly identified with probability converging to $1$ and the FWER is controlled at the $1-\gamma$ level. 
\begin{eqnarray*}
&&\lim_{r\to+\infty}\P\left(\forall i \in I,S(\widehat{\beta}_i^{\rm L}(\lambda,r))=s^0_i\right)=1,\\
&&\limsup_{r\to+\infty}\P\left(\exists i \notin I, \widehat{\beta}_i^{\rm L}(\lambda,r)\neq 0\right)\le  
1-\gamma,\\
&& \liminf_{r\to+\infty}\P\left(\exists i \notin I, \widehat{\beta}_i^{\rm L}(\lambda,r)\neq 0\right)\ge  
1-\bar{\gamma}.
\end{eqnarray*}
\end{description}
\end{prop}

\begin{Remark} The results in Proposition \ref{CI} are quite straightforward when $X$ is orthogonal ({\it i.e.} when $X'X=I$). Indeed, in this case the upper bound is simply the probability that zero components of $\beta$
are simultaneously estimated to $0$, namely $\P(\forall i\notin {\rm supp}(\beta), \hat{\beta}^{\rm L}_i(\lambda)=0)$
\end{Remark}

If $\varepsilon$ has a covariance matrix $\sigma^2Id_n$, one can obtain asymptotic results by fixing $\beta$ and letting $\sigma$ tend to $0$. However, unlike in our asymptotic setting described in Assumption \ref{assm1}, such asymptotic results are very poor and give no information about the FWER. Indeed, when $\sigma$ tends to $0$, the upper bound tends to $0$ or $1$, depending on whether the irrepresentable condition for $\beta$ holds or not.

Let us remind that the FWER is equal to $\P(\exists i \notin {\rm supp}(\beta), \widehat{\beta}^L_i\neq 0)$. According to Proposition \ref{CI}, when non-null components of $\beta$ are infinitely large, the FWER is controlled at level 
$1-\P\left(\left\|\zeta_{X,\lambda,{\rm sign}(\beta)}\right\|_\infty < 1\right)$. 
To our knowledge, it is  first theoretical result providing a formula for the FWER. 
Hereafter, let us provide some comments about the FWER control. 
\begin{itemize}
\item To provide a specific value for $\lambda$ allowing to control the FWER one needs to know the distribution of $\varepsilon$.
 For example, when the distribution of $\varepsilon$ is known and $\beta={\bf 0}$ one controls the FWER at level $\alpha$ by taking 
$\lambda$ as the $1-\alpha$ quantile of $\|X'\varepsilon\|_\infty$.  Let us point out that when 
$\varepsilon_1,\dots,\varepsilon_p$  are iid and the variance $\sigma^2$ of these components is unknown then $\sigma^2$ can be consistently estimate as explained in \cite{dicker}.
\item It is easier to control the FWER when $X$ is a random matrix whose distribution is symmetric and invariant by columns permutation  
 than when $X$ is a fixed design matrix. Indeed, when $X$ is random, the distribution of $\zeta_{X,\lambda,S(\beta)}$
just depends from the sparsity of $\beta$ and not on $S(\beta)$. 
\item Let $k\in \{1,\dots,p\}$ and let $U_k$ be the set $\{u \in \{-1,0,1\}^p\mid {\rm card}({\rm supp}(u))=k\}$ and let us assume
that $\beta \in U_k$. 
 When $X$ is a random matrix whose distribution is symmetric and invariant by columns permutation
by taking $\lambda_\alpha$ such that $\P(\|\zeta_{X,\lambda_\alpha,u_0}\|<1)=1-\alpha$ one controls asymptotically the FWER 
at level $\alpha$ (where $u_0:=(1,\dots,1,0,\dots,0)\in U_k$). Such a tuning parameter $\lambda_\alpha$ is easy to infer by Monte Carlo simulations. When $X$ is fixed, by taking $\lambda_\alpha$ as follows 
\begin{equation}\label{formulelambda}\frac{1}{{\rm card}(U_k)}\sum_{u\in U_k}\P(\|\zeta_{X,\lambda_\alpha,u}\|<1)=1-\alpha,\end{equation}
then the average value of the FWER with respect to $\beta\in U_k$   is $1-\alpha$ (where non-null components of $\beta$ are infinitely large). Again $\lambda_\alpha$ is easy to infer. 
In the numerical experiment, the tuning parameter was selected by solving equation (\ref{formulelambda}). 
\end{itemize}

\subsection{Proof of the Proposition \ref{CI}}
First, let us provide lemmas which are useful to prove both Proposition \ref{CI} and Theorem \ref{convergence}. Lemma \ref{lemmacvg2} partially proves  Proposition \ref{CI}. Indeed, according to this Lemma, when $(\beta^{(r)})_{r \in \N}$ is a sequence of $\R^p$  satisfying  assumptions \ref{assm1} then the following asymptotic result holds
$$\lim_{r\to +\infty}\P\left(\forall i \in {\rm supp}(s^0), S(\widehat{\beta}^{\rm L}_i(\lambda,r))=s^0_i\right)=1.$$
\begin{lemm}
\label{lemma1} 
Let $(\beta^{(r)})_{r \in \N}$ be a sequence of $\R^p$  satisfying the conditions {\bf 1)} and {\bf 2)} of  Assumption \ref{assm1}, let us assume that $s^0$ is identifiable with respect to the $\ell_1$ norm 
and let us set $u_r=\|\beta^{(r)}\|_1$ then
$$\lim_{r\to +\infty}\frac{\widehat{\beta}^{\rm L}(\varepsilon,r)-\beta^{(r)}}{u_r}=0.$$
\end{lemm}
{\bf Proof:} Because $\widehat{\beta}^{\rm L}(\varepsilon,r)$ is the LASSO estimator as defined in (\ref{LASSOpen}) then the following inequality occurs
$$
\frac{1}{2}\|Y-X\widehat{\beta}^{\rm L}(\varepsilon,r)\|^2_2+\lambda\|\widehat{\beta}^{\rm L}(\varepsilon,r)\|_1\le
\frac{1}{2}\|Y-X\beta^{(r)}\|^2_2+\lambda\|\beta^{(r)}\|_1. 
$$
Since $Y-X\beta^{(r)}=\varepsilon$  one may deduce the following inequalities
\begin{eqnarray}
\nonumber &&\lambda\|\widehat{\beta}^{\rm L}(\varepsilon,r)\|_1\le
\frac{1}{2}\|\varepsilon\|^2_2+\lambda\|\beta^{(r)}\|_1,\\
\label{ineq0} \Rightarrow&& \|\widehat{\beta}^{\rm L}(\varepsilon,r)/u_r\|_1
\le \frac{\|\varepsilon\|^2_2}{2\lambda u_r}+ 1. 
\end{eqnarray}
In addition, Cauchy-Schwarz inequality gives the following implications 
\begin{eqnarray}
\nonumber && \frac{1}{2}\|\varepsilon+X\beta^{(r)}-X\widehat{\beta}^{\rm L}(\varepsilon,r)\|^2_2+\lambda\|\widehat{\beta}^{\rm L}(\varepsilon,r)\|_1\le
\frac{1}{2}\|\varepsilon\|^2_2+\lambda\|\beta^{(r)}\|_1,\\ 
\nonumber \Rightarrow&& -\|\varepsilon\|_2\|X\beta^{(r)}-X\widehat{\beta}^{\rm L}(\varepsilon,r)\|_2+\frac{1}{2}
\|X\beta^{(r)}-X\widehat{\beta}^{\rm L}(\varepsilon,r)\|^2_2+\lambda\|\widehat{\beta}^{\rm L}(\varepsilon,r)\|_1\le
\lambda\|\beta^{(r)}\|_1,\\ 
\label{ineq1} \Rightarrow&&
-\frac{\|\varepsilon\|_2}{u_r}\left\|X\left(\frac{\widehat{\beta}^{\rm L}(\varepsilon,r)-\beta^{(r)}}{u_r}\right)\right\|_2+
\frac{1}{2}\left\|X\left(\frac{\widehat{\beta}^{\rm L}(\varepsilon,r)-\beta^{(r)}}{u_r}\right)\right\|^2_2+
\frac{\lambda}{u_r}\left\|\frac{\widehat{\beta}^{\rm L}(\varepsilon,r)}{u_r}\right\|_1\le\frac{\lambda}{u_r}.
\end{eqnarray}
Because $u_r$ tends to $+\infty$ then, according to (\ref{ineq0}), the sequence   $((\widehat{\beta}^{\rm L}(\varepsilon,r)-\beta^{(r)})/u_r)_{r\in \N^*}$ is bounded
since the following superior limit is finite
$$\limsup_{r \to +\infty}\left\|\frac{\widehat{\beta}^{\rm L}(\varepsilon,r)-\beta^{(r)}}{u_r}\right\|_1\le 2.$$ Consequently, to prove that 
$\lim_{r\to +\infty}(\widehat{\beta}^{\rm L}(\varepsilon,r)-\beta^{(r)})/u_r={\bf 0}$ 
it is sufficient to show that ${\bf 0}$ is the unique limit point of this sequence. 
Let $((\widehat{\beta}^{\rm L}(\varepsilon,\phi(r))-\beta^{(\phi(r))})/u_{\phi(r)})_{r\in \N^*}$ be a converging subsequence to $l$
(with $\phi: \N^* \to \N^*$  strictly increasing) and without loss of generality, let us assume 
$\lim_{r \to +\infty}\widehat{\beta}^{\rm L}(\varepsilon,\phi(r))/u_{\phi(r)}=v$ and 
$\lim_{r \to +\infty}\beta^{(\phi(r))}/u_{\phi(r)}=v'$ so that $l=v-v'$.   By (\ref{ineq0}) and (\ref{ineq1})
one may deduce that  
$$Xv=Xv'\text{ \rm and } \|v\|_1\le 1.$$
Since, whatever $r\ge 0$, we have $S(\beta^{(\phi(r))}/u_{\phi(r)})=s^0$ where $s^0$ is identifiable with respect to the $\ell_1$ norm
then, according to Proposition \ref{sign}, one may deduce that $\beta^{(\phi(r))}/u_{\phi(r)}$ is an unitary vector satisfying the identifiability condition. Consequently, $\|v'\|_1=1$ and $v'$ is identifiable with respect to the $\ell_1$ norm. Consequently,  
$v=v'$ and thus $l={\bf 0}$ is the unique limit point, which implies that 
$$\lim_{r\to +\infty} \frac{\widehat{\beta}^{\rm L}(\varepsilon,r)-\beta^{(r)}}{u_r}={\bf 0}.$$ 
$\hfill \square$\\\\
For the proof of  Lemma \ref{lemma1},  we have not used the third condition of  Assumption \ref{assm1}. This condition, under which the smallest non-null component of $\beta^{(r)}$ is not asymptotically infinitely smaller than $\|\beta^{(r)}\|_\infty$, is useful to prove  Lemma \ref{lemmacvg2}.
\begin{lemm}
\label{lemmacvg2}
Let $(\beta^{(r)})_{r \in \N}$ be a sequence of $\R^p$ satisfying  Assumption \ref{assm1}, then

$$\lim_{r\to +\infty}\P(\forall i \in {\rm supp}(s^0), S(\widehat{\beta}^{\rm L}_i(\lambda,r))=s^0_i)=1.$$
\end{lemm}
{\bf Proof:} Let $\varepsilon$ be a fixed vector in $\R^p$. According to the third condition of  Assumption \ref{assm1} we have 
$\min\{|\beta^{(r)}_i|, i \in {\rm supp}(s^0)\}/\|\beta^{(r)}\|_\infty\ge q>0$, consequently the following inequalities occur 
$$\forall i \in {\rm supp}(s^0), 
s^0_i\frac{\widehat{\beta}_i^{\rm L}(\varepsilon,\lambda,r)-\beta_i^{(r)}}{\|\beta^{(r)}\|_\infty}
= \frac{s^0_i\widehat{\beta}_i^{\rm L}(\varepsilon,\lambda,r)}{\|\beta^{(r)}\|_\infty}-\frac{|\beta_i^{(r)}|}{\|\beta^{(r)}\|_\infty}\le 
\frac{s^0_i\widehat{\beta}_i^{\rm L}(\varepsilon,\lambda,r)}{\|\beta^{(r)}\|_\infty}-q.$$
According to  Lemma \ref{lemma1}, the following inequality occurs
$$0=\liminf_{r\to +\infty}s^0_i\frac{\widehat{\beta}_i^{\rm L}(\varepsilon,\lambda,r)-\beta_i^{(r)}}{\|\beta^{(r)}\|_\infty}
\le \liminf_{r\to +\infty}\frac{s^0_i\widehat{\beta}_i^{\rm L}(\varepsilon,\lambda,r)}{\|\beta^{(r)}\|_\infty}-q. $$
Which implies that for $r$ large enough $s^0_i\widehat{\beta}_i^{\rm L}(\varepsilon,\lambda,r)>0$ and thus 
 $S(\widehat{\beta}_i^{\rm L}(\varepsilon,\lambda,r))=s^0_i$. When $\varepsilon$ is no longer fixed then,  for $i\in {\rm supp}(s^0)$, almost surely 
$S(\widehat{\beta}_i^{\rm L}(r))$ converges to $s^0_i$ and  consequently 
$$\lim_{r\to +\infty}\P\left(\forall i \in {\rm supp}(s^0), S(\widehat{\beta}^{\rm L}_i(\lambda,r))=s^0_i\right)=1.$$
$\hfill \square$\\\\
{\bf Proof of  Proposition \ref{CI}:}\\
 Let us remind that the vector $\widehat{\beta}^{\rm L}(\lambda)$ is the LASSO estimator if and only if the  following two inequalities occur simultaneously.
\begin{eqnarray}
\label{egalite}X_A'(Y-X\widehat{\beta}^{\rm L}(\lambda))&=&\lambda S(\widehat{\beta}^{\rm L}_A(\lambda)), \text{ \rm where }
A={\rm supp}(\widehat{\beta}^{\rm L}(\lambda)),\\
\label{inegalite}\|X_{\overline{A}}'(Y-X\widehat{\beta}^{\rm L}(\lambda))\|_\infty&\le& \lambda.
\end{eqnarray} 
{\bf Sharpness of the upper bound)} 
Since the upper bound  depends only on $s^0$ and not on how large  the non-null components $\beta^{(r)}$ are then  
$$\limsup_{r\to +\infty}\P\left(S(\widehat{\beta}^{\rm L}(\lambda,r))=s^0\right)\le \P\left(\left\|\zeta_{X,\lambda,s^0}\right\|_\infty \le 1\right).$$
Finally, it must be proven that $\liminf_{r\to +\infty}\P\left(S(\widehat{\beta}^{\rm L}(\lambda,r))=s^0\right)\ge  
\P\left(\left\|\zeta_{X,\lambda,s^0}\right\|_\infty < 1\right)$.
Let us remind that $I={\rm supp}(s^0)$ and
let us assume that the following events hold simultaneously
\begin{equation}\label{keystone}
X_I'(Y-X\widehat{\beta}^{\rm L}(\lambda))=\lambda s^0_I \text{ \rm and }\underbrace{\left\|X_{\overline{I}}'X_I(X_I'X_I)^{-1}\lambda 
s^0_I+X_{\overline{I}}'\left(Id-X_I(X_I'X_I)^{-1}X_I'\right)\varepsilon\right\|_\infty < \lambda
}_{=\left\|\zeta_{X,\lambda,s^0}\right\|_\infty < 1}. 
\end{equation}
We aim to show that the inequalities given above imply that $\widehat{\beta}^{\rm L}_{\overline{I}}(\lambda)={\bf 0}$. 
For convenience, let us set $H$ be the projection matrix 
$H:=X_I(X_I'X_I)^{-1}X_I'$. When (\ref{keystone}) occurs then the following inequalities holds

\begin{eqnarray}
\nonumber \left\|X_{\overline{I}}'H(Y-X\widehat{\beta}^{\rm L}(\lambda))+
X_{\overline{I}}'\left(Id-H\right)\varepsilon\right\|_\infty &<& \lambda,\\
\nonumber \left\|X_{\overline{I}}'\left(H(Y-X\widehat{\beta}^{\rm L}(\lambda))+
(Id-H)\varepsilon\right)\right\|_\infty &<& \lambda,\\
\label{ineg} \left\|X_{\overline{I}}'\left(Y-X\widehat{\beta}^{\rm L}(\lambda)+X_{\overline{I}}\widehat{\beta}^{\rm L}_{\overline{I}}(\lambda)-
HX_{\overline{I}}\widehat{\beta}^{\rm L}_{\overline{I}}(\lambda)\right)\right\|_\infty &<& \lambda. 
\end{eqnarray}
Inequality (\ref{ineg}) comes from the following two identities 
\begin{eqnarray*}HY&=&H(X\beta^{(r)})+H\varepsilon=H(X_I\beta_I^{(r)})+H\varepsilon=X_I\beta_I^{(r)}+H\varepsilon
=X(\beta^{(r)})+H\varepsilon \text{ \rm and},\\ 
HX\widehat{\beta}^{\rm L}(\lambda)&=&HX_I\widehat{\beta}^{\rm L}_I(\lambda)
+HX_{\overline{I}}\widehat{\beta}^{\rm L}_{\overline{I}}(\lambda)=
X_I\widehat{\beta}^{\rm L}_I(\lambda)
+HX_{\overline{I}}\widehat{\beta}^{\rm L}_{\overline{I}}(\lambda)=X\widehat{\beta}^{\rm L}(\lambda)
-X_{\overline{I}}\widehat{\beta}^{\rm L}_{\overline{I}}(\lambda)
+HX_{\overline{I}}\widehat{\beta}^{\rm L}_{\overline{I}}(\lambda).
\end{eqnarray*}
Let $v$  be the vector $v:=X_{\overline{I}}'\left(Y-X\widehat{\beta}^{\rm L}(\lambda)+X_{\overline{I}}\widehat{\beta}^{\rm L}_{\overline{I}}(\lambda)-HX_{\overline{I}}\widehat{\beta}^{\rm L}_{\overline{I}}(\lambda)\right)$. We are going to see that   inequality (\ref{ineg}) implies that 
$\widehat{\beta}^{\rm L}_{\overline{I}}(\lambda)={\bf 0}$. Let us assume that 
$\widehat{\beta}^{\rm L}_{\overline{I}}(\lambda)\neq {\bf 0}$
 then, on the one hand, the following inequality occurs 
\begin{equation}\label{contradiction1}\widehat{\beta}^{\rm L}_{\overline{I}}(\lambda)'v\le \|\widehat{\beta}^{\rm L}_{\overline{I}}(\lambda)\|_1\|v\|_\infty<
\lambda\|\widehat{\beta}^{\rm L}_{\overline{I}}(\lambda)\|_1.\end{equation}
According to (\ref{egalite}) the identity $\widehat{\beta}^{\rm L}_i(\lambda)X_i'(Y-X\widehat{\beta}^{\rm L}(\lambda))=
\lambda|\widehat{\beta}^{\rm L}_i(\lambda)|$ occurs. Consequently, on the other hand, the following inequalities hold
\begin{eqnarray}
\nonumber \widehat{\beta}^{\rm L}_{\overline{I}}(\lambda)'v&=&\widehat{\beta}^{\rm L}_{\overline{I}}(\lambda)'X_{\overline{I}}'
\left(Y-X\widehat{\beta}^{\rm L}(\lambda)+X_{\overline{I}}\widehat{\beta}^{\rm L}_{\overline{I}}(\lambda)-HX_{\overline{I}}
\widehat{\beta}^{\rm L}_{\overline{I}}(\lambda)\right),\\
\nonumber &=&\lambda\|\widehat{\beta}^{\rm L}_{\overline{I}}(\lambda)\|_1+\widehat{\beta}^{\rm L}_{\overline{I}}(\lambda)'X_{\overline{I}}'(Id-H)
X_{\overline{I}}\widehat{\beta}^{\rm L}_{\overline{I}}(\lambda),\\
\label{contradiction2}&\ge& \lambda \|\widehat{\beta}^L_{\overline{I}}(\lambda)\|_1.
\end{eqnarray}
The last inequality occurs because the projection matrix  $Id-H$ is  positive semi-definite. Inequalities (\ref{contradiction1}) and 
(\ref{contradiction2}) provide a contradiction 
which implies that $\widehat{\beta}^{\rm L}_{\overline{I}}(\lambda)={\bf 0}$. \\
According to (\ref{egalite}), the following implication holds
$$ S(\widehat{\beta}^{\rm L}_I(\lambda,r))=s^0_I
\Rightarrow X_I'(Y-X\widehat{\beta}^{\rm L}(\lambda,r))=\lambda s^0_I.$$
Because $s^0$ is identifiable with respect to the $\ell_1$ norm then, according to  Lemma \ref{lemmacvg2},
the following convergence in probability occurs
\begin{equation} \label{power}\lim_{r\to +\infty}\P( S(\widehat{\beta}^{\rm L}_I(\lambda,r))=s^0_I)=
\lim_{r\to +\infty}\P(X_I'(Y-X\widehat{\beta}^{\rm L}(\lambda,r))=\lambda s^0_I)=1.\end{equation}
Using this asymptotic result and since when (\ref{keystone}) occurs then $\widehat{\beta}^{\rm L}_{\overline{I}}(\lambda,r)={\bf 0}$, one  may deduce the following inequalities
\begin{eqnarray*}
\liminf_{r\to +\infty}\P\left(S(\widehat{\beta}^{\rm L}(\lambda,r))=s^0\right)&=&
\liminf_{r\to +\infty}\P\left( S(\widehat{\beta}^{\rm L}_I(\lambda,r))=s^0_I \text{ \rm and }
\widehat{\beta}^{\rm L}_{\overline{I}}(\lambda,r)={\bf 0}\right),\\
&=& \liminf_{r\to +\infty}\P(\widehat{\beta}^{\rm L}_{\overline{I}}(\lambda,r)={\bf 0}),\\
&\ge & \liminf_{r\to +\infty}\P\left(X_I'(Y-X\widehat{\beta}^{\rm L}(\lambda,r))=\lambda s^0_I \text{ \rm and }
\left\|\zeta_{X,\lambda,s^0}\right\|_\infty < 1\right),\\
&\ge& \liminf_{r\to +\infty}\P\left(\left\|\zeta_{X,\lambda,s^0}\right\|_\infty < 1\right).
\end{eqnarray*}
{\bf Asymptotic full power and asymptotic control of the FWER)} 
According to (\ref{power}), asymptotically the power is equal to $1$, namely $\lim_{r\to +\infty}\P(\forall i \in I, S(\widehat{\beta}^{\rm L}_i(\lambda,r))=s^0_i)=1$. 
Now let us prove that the FWER is controlled asymptotically. Let us remind that 
$\P\left(\left\|\zeta_{X,\lambda,s^0}\right\|_\infty < 1\right)=\gamma$ and $\P\left(\left\|\zeta_{X,\lambda,s^0}\right\|_\infty \le 1\right)=\bar{\gamma}$.
Using asymptotic results given above one may deduce the following 
inequalities. 
\begin{eqnarray}
\nonumber \bar{\gamma} &\ge &\limsup_{r\to +\infty}\P(S(\widehat{\beta}^L(\lambda,r))=s^0),\\
\nonumber &\ge& \limsup_{r\to +\infty}\P\left(\forall i \in I, S(\widehat{\beta}^L_i(\lambda,r))=s^0_i
\text{ \rm and }\forall i\notin I, \widehat{\beta}^L_i(\lambda,r)=0\right),\\
\label{lim1}&\ge & \limsup_{r\to +\infty}\P(\forall i\notin I, \widehat{\beta}^L_i(\lambda,r)=0).
\end{eqnarray}
The last inequality comes from (\ref{power}). Similarly, we have 
\begin{equation} \label{lim2}
\gamma\le \liminf_{r\to +\infty}
\P(\forall i\notin I, \widehat{\beta}^L_i(\lambda,r)=0).\end{equation}
Consequently, by taking the complement to $1$ of the inequalities given in (\ref{lim1}) and (\ref{lim2}), one may deduce that
$$\liminf_{r\to +\infty}
\P(\exists i\notin I, \widehat{\beta}^L_i(\lambda,r)\neq 0)\ge 1-\bar{\gamma} \text{ \rm and }
\limsup_{r\to +\infty}\P(\exists i\notin I, \widehat{\beta}^L_i(\lambda,r)\neq 0)\le 1-\gamma.$$
$\hfill \square$

\subsection*{\bf Proof of  Theorem \ref{convergence}}
 Lemma \ref{lemma3} provides the same result for BPDN as does  Lemma \ref{lemma1} for LASSO.
These both lemmas are the keystones to prove  Theorem \ref{convergence}.

\begin{lemm}
\label{lemma3} 
Let $(\beta^{(r)})_{r \in \N}$ be a sequence of $\R^p$  satisfying  conditions {\bf 1)} and {\bf 2)} of  Assumption \ref{assm1}, 
let us assume that $s^0$ is identifiable with respect to the $\ell_1$ norm 
and let set $u_r=\|\beta^{(r)}\|_1$ then
$$\lim_{r\to +\infty}\frac{\widehat{\beta}^{\rm BPDN}(\varepsilon,r)-\beta^{(r)}}{u_r}=0.$$
\end{lemm}
{\bf Proof:} 
Let us define $u(\varepsilon)\in \R^p$ as follows 
$$u(\varepsilon):=\underset{b \in \R^p}{\text{\rm argmin }}\|b\|_1 
\text{ \rm subject to }Xb=\varepsilon.$$
Because  $X(u(\varepsilon))=\varepsilon$, we have $Y(\varepsilon)=X(\beta^{(r)}+u(\varepsilon))$ and because $\widehat{\beta}^{\rm BPDN}(\varepsilon,r)$ is an admissible point of (\ref{BPDN}),  one deduces the  following inequality 
\begin{equation}
\label{eq1}\left\|\frac{1}{u_r}X\widehat{\beta}^{\rm BPDN}(\varepsilon,r)-\frac{1}{u_r}X\beta^{(r)}\right\|_2\le 
\left\|\frac{1}{u_r}X\widehat{\beta}^{\rm BPDN}(\varepsilon,r)-\frac{1}{u_r}Y\right\|_2+
\left\|\frac{1}{u_r}Y-\frac{1}{u_r}X \beta^{(r)}\right\|_2 \le 
\frac{\sqrt{R}}{u_r}+\frac{\|Xu(\varepsilon)\|_2}{u_r}.
\end{equation}
 Because $\beta^{(r)}+u(\varepsilon)$ is an admissible point of  problem (\ref{BPDN}) and  
because $\widehat{\beta}^{\rm BPDN}(\varepsilon,r)$ is the minimizer of (\ref{BPDN}), one may deduce that the following inequalities hold
\begin{equation}
\label{eq2}\frac{1}{u_r}\|\widehat{\beta}^{\rm BPDN}(\varepsilon,r)\|_1\le 
\frac{1}{u_r}\|\beta^{(r)}+u(\varepsilon)\|_1\le 1+\frac{\|u(\varepsilon)\|_1}{u_r}.
\end{equation}
Because $u_r$ tends to $+\infty$ then, according to (\ref{eq2}), the sequence   $((\widehat{\beta}^{\rm L}(\varepsilon,r)-\beta^{(r)})/u_r)_{r\in \N^*}$ is bounded
since the following superior limit is finite
$$\limsup_{r \to +\infty}\left\|\frac{\widehat{\beta}^{\rm BPDN}(\varepsilon,r)-\beta^{(r)}}{u_r}\right\|_1\le 2.$$ Consequently, to prove that 
$\lim_{r\to +\infty}(\widehat{\beta}^{\rm BPDN}(\varepsilon,r)-\beta^{(r)})/u_r={\bf 0}$ 
it is sufficient to show that ${\bf 0}$ is the unique limit point of this sequence. 
Let $((\widehat{\beta}^{\rm L}(\varepsilon,\phi(r))-\beta^{(\phi(r))})/u_{\phi(r)})_{r\in \N^*}$ be a converging subsequence to $l$
(with $\phi: \N^* \to \N^*$  strictly increasing) and without loss of generality, let us assume 
$\lim_{r \to +\infty}\widehat{\beta}^{\rm BPDN}(\varepsilon,\phi(r))/u_{\phi(r)}=v$ and 
$\lim_{r \to +\infty}b^{(\phi(r))}/u_{\phi(r)}=v'$ so that $l=v-v'$.   By (\ref{eq1}) and (\ref{eq2})
one may deduce that  
$$Xv=Xv'\text{ \rm and } \|v\|_1\le 1.$$
Since, whatever $r\ge 0$, we have $S(\beta^{(\phi(r))}/u_{\phi(r)})=s^0$ where $s^0$ is identifiable with respect to the $\ell_1$ norm
then, according to  Proposition \ref{sign}, one may deduce that $\beta^{(\phi(r))}/u_{\phi(r)}$ is an unitary vector satisfying the identifiability condition. Consequently, $\|v'\|_1=1$ and $v'$ is identifiable with respect to the $\ell_1$ norm. Consequently,  
$v=v'$ and thus $l={\bf 0}$ is the unique limit point, which implies that 
$$\lim_{r\to +\infty} \frac{\widehat{\beta}^{\rm BPDN}(\varepsilon,r)-\beta^{(r)}}{u_r}={\bf 0}.$$ 
$\hfill \square$\\\\
 Lemma \ref{idtf} is useful to prove in Theorem \ref{convergence}  that when $s^0$ is not identifiable then
 sign estimator derived from thresholded LASSO cannot recover $s^0$.  
\begin{lemm}
\label{idtf}
Let $X$ be a matrix in general position, then 
the random vector $\widehat{\beta}$ is identifiable with respect to $X$ and the $\ell_1$ norm.
\end{lemm}
{\bf Proof:} Let us remind that when $X$ is in general position then the minimizer $\widehat{\beta}$ is unique. 
 Let us assume that $\widehat{\beta}$ is not identifiable with respect to $X$ and the $\ell_1$ norm, then 
there exists $b\in \R^p$ such that $Xb=X\widehat{\beta}$ and $\|b\|_1\le \|\widehat{\beta}\|_1$. 
Consequently, for LASSO, one may deduce that 
$$\|Y-Xb\|^2+\lambda\|b\|_1\le \|Y-X\widehat{\beta}^{\rm L}\|^2+\lambda\|\widehat{\beta}^{\rm L}\|_1.$$  
This inequality  contradicts  
$\widehat{\beta}^{\rm L}$ as the unique minimizer of (\ref{LASSOpen}). Similarly, when $\widehat{\beta}^{\rm BPDN}$ is not identifiable with respect to the $\ell_1$ norm then 
$\widehat{\beta}^{\rm BPDN}$ is not the unique minimizer of (\ref{BPDN}), which provides a contradiction. $\hfill \square$\\\\
For the proofs of  Theorem \ref{convergence} and the proof of   Proposition \ref{sign} we need to 
introduce the following inequality which characterizes the identifiability condition \cite{daubechies2010}. 
A vector  $b\in \R^p$ is identifiable with respect to $X$ and the $\ell_1$ norm if and only if the following inequality holds 
\begin{equation}\label{daubechies caracterisation}
\forall h \in \ker(X)\setminus\{{\bf 0}\}, \left|\sum_{i \in {\rm supp}(b)}S(b)h_i\right|< 
\sum_{i \notin {\rm supp}(b)}|h_i|. 
\end{equation}
\\\\
{\bf Proof of  Theorem \ref{convergence}:}\\
{\bf Necessary condition:} Let us assume that $S(\beta)$ is not identifiable with respect to the $\ell_1$ norm. Let us show that when the following events hold
\begin{equation} \label{event}{\rm supp}^-(\beta)\subset {\rm supp}^-( \widehat{\beta}(\varepsilon)) \text{ \rm and }{\rm supp}^+(\beta)\subset 
{\rm supp}^+( \widehat{\beta}(\varepsilon)),\end{equation}
 then  inequality (\ref{daubechies caracterisation}) occurs which contradicts that $S(\beta)$ is not identifiable 
with respect to the $\ell_1$ norm. Let $h\in \ker(X)\setminus\{{\bf 0}\}$. On the one hand, when (\ref{event}) occurs, we have
\begin{eqnarray*}\left|\sum_{i \in {\rm supp}(\beta)}S(\beta_i)h_i\right|&=&\left| -\sum_{{\rm supp}^-(\beta)}h_i+\sum_{{\rm supp}^+(\beta)}h_i \right|,\\
&=& \left| -\sum_{i \in {\rm supp}^-( \widehat{\beta}(\varepsilon))}h_i+\sum_{i \in {\rm supp}^-(\widehat{\beta}(\varepsilon))\setminus {\rm supp}^-(\beta) }h_i  +\sum_{i \in {\rm supp}^+( \widehat{\beta}(\varepsilon))}h_i-\sum_{i \in {\rm supp}^+(\widehat{\beta}(\varepsilon))\setminus {\rm supp}^+(\beta) }h_i  \right|,\\
&\le& 
\left| -\sum_{i \in {\rm supp}^-( \widehat{\beta}(\varepsilon))}h_i+\sum_{i \in {\rm supp}^+( \widehat{\beta}(\varepsilon))}h_i \right|+
\sum_{i \in {\rm supp}(\widehat{\beta}(\varepsilon))\setminus {\rm supp}(\beta) }|h_i|.
\end{eqnarray*}
On the other hand, according to  Lemma \ref{idtf}, $\widehat{\beta}(\varepsilon)$ is identifiable with respect to the $\ell_1$ norm then  
(\ref{daubechies caracterisation}) occurs implying  the following inequality
\begin{multline*}\left| -\sum_{i \in {\rm supp}^-( \widehat{\beta}(\varepsilon))}h_i+\sum_{i \in {\rm supp}^+( \widehat{\beta}(\varepsilon))}h_i \right|+
\sum_{i \in {\rm supp}(\widehat{\beta}(\varepsilon))\setminus {\rm supp}(\beta) }|h_i|\\<\sum_{i \notin {\rm supp}(\widehat{\beta}(\varepsilon))}|h_i|+
\sum_{i \in {\rm supp}(\widehat{\beta}(\varepsilon))\setminus {\rm supp}(\beta) }|h_i|= \sum_{i \notin {\rm supp}(\beta)}|h_i|.
\end{multline*}
Consequently the following inequality holds
$$\forall h \in \ker(X)\setminus\{{\bf 0}\}, \left|\sum_{i \in {\rm supp}(\beta)}S(\beta_i)h_i\right|
<\sum_{i \notin {\rm supp}(\beta)}|h_i|,$$ 
which, according to (\ref{daubechies caracterisation}), contradicts that $S(\beta)$ is not identifiable with respect to the $\ell_1$ norm.\\\\ 
{\bf Sufficient condition:} Let us remind that according to  condition {\bf 3)} of  Assumption \ref{assm1}  the  following inequality holds
$$\forall r\in \N, \frac{\min\{|\beta^{(r)}_i|, i \in {\rm supp}(s^0)\}}{\|\beta^{(r)}\|_\infty}\ge q>0.$$
According to  Lemmas \ref{lemma1} and \ref{lemma3}, when $s^0$ is identifiable with respect to the $\ell_1$ norm then 
$$\lim_{r\to +\infty}\frac{\widehat{\beta}(\varepsilon,r)-\beta^{(r)}}{\|\beta^{(r)}\|_\infty}=0.$$
Therefore, there exists 
$r_0(\varepsilon)\ge 0$ such that
$$
\forall r\ge  r_0(\varepsilon), \left\|\frac{\widehat{\beta}(\varepsilon,r)-\beta^{(r)}}{\|\beta^{(r)}\|_\infty}\right\|_\infty < q/2 \Leftrightarrow
\forall i\in \{1,\dots,p\}, \forall r\ge  r_0(\varepsilon), \left|\frac{\widehat{\beta}_i(\varepsilon,r)-\beta_i^{(r)}}{\|\beta^{(r)}\|_\infty}\right| < q/2.
$$
Consequently, when $r\ge r_0(\varepsilon)$, whatever $i\notin {\rm supp}(s^0)$ (thus when $\beta_i^{(r)}=0$) the following inequalities hold
\begin{eqnarray*}
&& \forall i \notin {\rm supp}(s^0), \left|\frac{\widehat{\beta}_i(\varepsilon,r)}{\|\beta^{(r)}\|_\infty}\right| < q/2,\\
&\Rightarrow& -\|\beta^{(r)}\|_\infty q/2 < \min_{i \notin {\rm supp}(s^0)}\left\{\widehat{\beta}_i(\varepsilon,r)\right\}
\le \max_{i \notin {\rm supp}(s^0)}\left\{\widehat{\beta}_i(\varepsilon,r)\right\} < \|\beta^{(r)}\|_\infty q/2. 
\end{eqnarray*}
Whatever $i\in {\rm supp}^+(s^0)$ (thus when $\beta_i^{(r)}>0)$ the following inequalities hold
\begin{eqnarray*}
&& \forall i \in {\rm supp}^+(s^0),  \frac{\widehat{\beta}_i(\varepsilon,r)}{\|\beta^{(r)}\|_\infty}\ge  -\left|\frac{\widehat{\beta}_i(\varepsilon,r)-\beta_i^{(r)}}{\|\beta^{(r)}\|_\infty}\right|+\frac{\beta_i^{(r)}}{\|\beta^{(r)}\|_\infty},\\
&\Rightarrow&  \min_{i\in {\rm supp}^+(s^0)}\left\{\frac{\widehat{\beta}_i(\varepsilon,r)}{\|\beta^{(r)}\|_\infty}\right\}> -q/2+q
=q/2, \\
&\Rightarrow& \min_{i\in {\rm supp}^+(s^0)}\left\{\widehat{\beta}_i(\varepsilon,r)\right\}> \|\beta^{(r)}\|_\infty q/2.  
\end{eqnarray*}
Whatever $i\in {\rm supp}^-(s^0)$ (thus when $\beta_i^{(r)}<0)$ the following inequalities hold
\begin{eqnarray*}
&& \forall i \in {\rm supp}^+(s^0),  \frac{\widehat{\beta}_i(\varepsilon,r)}{\|\beta^{(r)}\|_\infty}\le  \left|\frac{\widehat{\beta}_i(\varepsilon,r)-\beta_i^{(r)}}{\|\beta^{(r)}\|_\infty}\right|+\frac{\beta_i^{(r)}}{\|\beta^{(r)}\|_\infty},\\
&\Rightarrow&  \max_{i\in {\rm supp}^-(s^0)}\left\{\frac{\widehat{\beta}_i(\varepsilon,r)}{\|\beta^{(r)}\|_\infty}\right\}< q/2-q
=-q/2, \\
&\Rightarrow& \max_{i\in {\rm supp}^-(s^0)}\left\{\widehat{\beta}_i(\varepsilon,r)\right\}< -\|\beta^{(r)}\|_\infty q/2.  
\end{eqnarray*}
Finally, when $r\ge r_0(\varepsilon)$ we have 
\begin{itemize} 
\item[i)] $$ {\rm supp}^-(s^0)\subset {\rm supp}^-(\widehat{\beta}_i(\varepsilon,r)) \text{ \rm and } 
{\rm supp}^+(s^0)\subset {\rm supp}^+(\widehat{\beta}_i(\varepsilon,r)). $$
\item[ii)]
$$\max_{i\in {\rm supp}^-(s^0)}\left\{\widehat{\beta}_i(\varepsilon,r)\right\}<\min_{i\notin {\rm supp}(s^0)}\left\{\widehat{\beta}_i(\varepsilon,r)\right\}\le \max_{i\notin {\rm supp}(s^0)}\left\{\widehat{\beta}_i(\varepsilon,r)\right\}< \min_{i\in {\rm supp}^+(s^0)}\left\{\widehat{\beta}_i(\varepsilon,r)\right\}.$$
\end{itemize}
These achieve  the proof of the sufficient condition. 
$\hfill \square$

\subsection*{Proof of propositions}
The  proof  of  Proposition \ref{CI_implique_NSP}, provided below, is the one reported in the PhD manuscript of Tardivel \cite{tardivelphd}. \\\\
{\bf Proof of  Proposition \ref{CI_implique_NSP}:}
From Daubechies et al. \cite{daubechies2010}, $\beta$ is a 
parameter having a minimal $\ell_1$ norm, namely $X\beta=X\gamma \Rightarrow\|\gamma\|_1\ge\|\beta\|_1$ holds if and only if 
the following inequality occurs
\begin{equation}
\label{daubechies caracterisation1}
\forall h\in \ker(X), \left|\sum_{i \in I}S(\beta_i)h_i\right|\le \sum_{i \notin I}|h_i|.
\end{equation}
We are going to show that when the irrepresentable condition holds for $\beta$ then the inequality   (\ref{daubechies caracterisation}) holds.

Let $h\in \ker(X)$ and let us remind that $h_I$ and $h_{\overline{I}}$ denote respectively vectors $(h_i)_{i\in I}$ and 
$(h_i)_{i\notin I}$. Then the following equality holds
$$\sum_{i \in I}S(\beta_i)h_i=h_I' S(\beta_I)= h_I'X_{I}'X_I(X_I'X_I)^{-1}S(\beta_I).$$
Because ${\bf 0}=Xh=X_I h_I +X_{\overline{I}} h_{\overline{I}}$, one may deduce the following inequalities 
\begin{eqnarray}
\nonumber|h_I' S(\beta_I)|&=& \left|h_{\overline{I}}'X_{\overline{I}}'X_I(X_I'X_I)^{-1}S(\beta_I)\right|,\\
\label{inegalite final} &\le & \|h_{\overline{I}}\|_1\|X_{\overline{I}}'X_I(X_I'X_I)^{-1}S(\beta_I)\|_\infty.
\end{eqnarray}
Consequently, when the irrepresentable condition  holds for $\beta$, namely when $\|X_{\overline{I}}' X_I(X_I'X_I)^{-1}S(b_I^*)\|_\infty\le 1$, then the inequality (\ref{inegalite final}) gives $|h_I' S(\beta_I)|\le \|h_{\overline{I}}\|_1$. 
Thus, by the equivalence given in (\ref{daubechies caracterisation1}), $\beta$ is a solution of the following  basis pursuit problem 
$$\text{\rm minimize } \|\gamma\|_1 \text{ \rm subject to }X\gamma=X\beta$$
Because $X$ is in general position the previous optimisation problem has a unique solution (see, {\it e.g.}, Proposition 1 in Appendix)
thus $X\beta=X\gamma$ and $\gamma \neq \beta$ implies that  $\|\gamma\|_1>\|\beta\|_1$, namely $\beta$ is identifiable with respect to the $\ell_1$ norm.
$\hfill \square$\\\\
Let us notice that when the inequality in the irrepresentable condition is strict,  Theorem \ref{CI_implique_NSP} remains true without  assuming  that $X$ is in general position.\\\\
{\bf Proof of  Proposition \ref{sign}:}
Because $b$ is identifiable with respect to the $\ell_1$ norm and because $S(\tilde{b})=S(b)$ implies ${\rm supp}(\tilde{b})={\rm supp}(b)$, then the following inequality holds 
$$\forall h \in \ker(X)\setminus\{{\bf 0}\}, \left|\sum_{i \in {\rm supp}(\tilde{b})}S(\tilde{b}_i)h_i\right|< \sum_{i \notin {\rm supp}(\tilde{b})}|h_i|. $$
Consequently, according to (\ref{daubechies caracterisation}),   parameter $\tilde{b}$ is identifiable with respect to the $\ell_1$ norm. $\hfill \square$

\subsection{Comparisons of  conditions for sign recovery and support recovery} 
Let $X=(X_1|\dots|X_p)\in \R^{n\times p}$ and 
$\beta \in \R^p$. 
\begin{itemize}
    \item When $\|X_1\|_2=\dots=\|X_p\|_2=1$, we 
say that $\beta$ satisfies the mutual coherence condition once the following inequalities occurs 
$${\rm card}({\rm supp}(\beta))<\frac{1+1/M(X)}{2}, \text{ \rm where }M(X)=\max_{i\neq j}|X_i'X_j|.$$
    \item We say that $\beta$ satisfies the irrepresentability condition once 
$$\|X_{\overline I}'X_I(X_I'X_I)^{-1}{\rm sign}(\beta_I)\|_\infty<1, \text{ \rm where }I={\rm supp}(\beta).$$
Moreover, we say that the irrepresentability condition
uniformly hold on the support $I$ of $\beta$ once 
$$\forall \theta \in \{-1,1\}^{|I|}, \,\|X_{\overline I}'X_I(X_I'X_I)^{-1}\theta\|_\infty<1.$$
    \item We say that $\beta$ satisfies the stable nullspace property once 
$$\forall h\in \ker(X)\setminus\{ 0\},  \sum_{i\in {\rm supp}(\beta)}|h_i|<\sum_{i\notin {\rm supp}(\beta)}|h_i|.$$
    \item  We remind that $\beta$ is identifiable with respect to $X$ and the $\ell_1$ norm once 
$$\forall h\in \ker(X)\setminus\{ 0\},  \left|\sum_{i\in {\rm supp}(\beta)}{\rm sign}(\beta_i)h_i\right|<\sum_{i\notin {\rm supp}(\beta)}|h_i|$$
\end{itemize}
Table \ref{tab:my_label2} summarizes comparisons between above conditions
\begin{table}[h]
    \centering
    \begin{tabular}{c c c c c c c c}
        \text{\rm Mutual coherence}  &  $\overset{\cite{vandegeer2008}}\Longrightarrow$   & \text{\rm Irrepresentability} & &  & & \\
$\Downarrow$\cite{gribonval} & & $\Downarrow$ & & \text{ \rm Uniform irrepresentability} & $\overset{\cite{descloux}}{\Longrightarrow}$ & 
\text{ \rm Stable nullspace}\\
\text{ \rm Stable nullspace} & $\Longrightarrow$ & {\rm Identifiability}& & & &
    \end{tabular}
    \caption{This figure provides implication between above conditions. The implication  ``stable nullspace  $\Rightarrow$ identifiability''  is straightforward  and the implication ``Irrepresentability $\Rightarrow$ Identifiability'' is given in Proposition 1. Other implications are proved in \cite{descloux,gribonval,vandegeer2008}. Note that
    an exhaustive scheme is given in \cite{buhlmannlivre,vandegeer2008}.}
    \label{tab:my_label2}
\end{table}

One may observe that the mutual coherence  is a very strong condition once two columns of $X$ are almost equal (namely when $M(X)$ is close to $1$). 
As an example, $M(X)=0.99$ when $X$ is the matrix from the riboflavin dataset and thus the mutual coherence condition is extremely strong. Nevertheless, 
when $\beta$ has only one non null component, the mutual coherence condition holds and thus $\beta$ satisfies both the irrepresentability and identifiability conditions as illustrated on Figure 5.

\begin{figure}[htbp]
\centering 
	\includegraphics[scale=0.5]{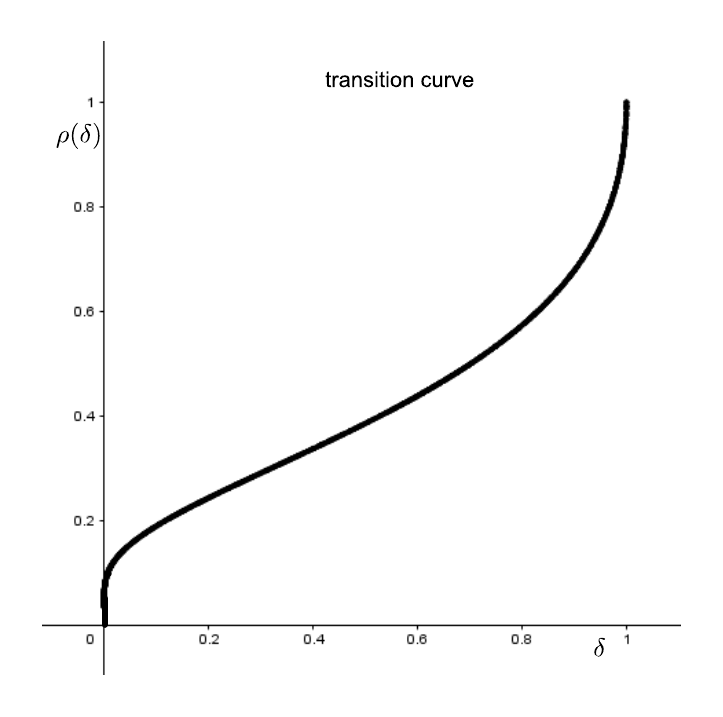}
	\caption{ This figure illustrates the transition curve
	%of the function $\rho: (0,1)\to (0,1)$ 
	%so called  transition curve 
	of Donoho and Tanner \cite{donoho_tanner}.}
	\label{curve_phase_transition}
\end{figure}

Finally when $X$ is a $n\times p$ standard Gaussian matrix and $n<p$ are both very large then $\beta$ is identifiable with respect to $X$ and the $\ell_1$ norm with a very large probability (resp. very small probability) once  
${\rm card}({\rm supp}(\beta))<n\rho(n/p)$ (resp. once ${\rm card}({\rm supp}(\beta))>n\rho(n/p)$), where $\rho(\cdot)$ is the transition curve of Donoho and Tanner \cite{donoho_tanner}, presented in Figure \ref{curve_phase_transition}.

\section*{Supplementary material}
We have already said that  when $X$ is in general position the minimizer of problem (\ref{LASSOpen}) (resp. problem (\ref{BPDN})) is unique. 
Concerning LASSO, a sketch of proof given in Tibshirani \cite{tibshiraniryan} shows 
the uniqueness of the LASSO estimator  when $X$ is in general position. In order to provide a self-contained article, we show that  when $X$ is in general position, the minimizer of  problem (\ref{BPDN}) is unique when $R=0$ as well as when $R>0$. We have already stressed that when $\beta$ is identifiable with respect to the $\ell_1$ norm then $\beta$ is sparse. We show that when the identifiability holds for $\beta$ then the family $(X_i)_{i\in {\rm supp}(\beta)}$ is linearly independent and thus the number of  components of $\beta$ equal to 0 is larger than $p-n$. Finally, a proof that the stable nullspace property implies the identifiability condition is given

\bibliographystyle{plain}
\bibliography{biblio_clean}
\end{document}

\doublespace
\title{Supplementary material}
\author{Patrick J.C. Tardivel$^{a}\footnote{Corresponding author: tardivel@math.uni.wroc.pl}\,$ and 
Ma\l{}gorzata Bogdan$^{a,b}$,\\
\small{ $^{a}$  Institute of Mathematics, University of Wroc{\l}aw, Wroc\l{}aw, Poland }\\
\small{ $^{b}$  Department of Statistics, Lund University, Lund, Sweden}}
\date{}
\maketitle
\section*{General position condition}
Let $M_1,\dots,M_k\in \R^p$ and let $\text{\rm Aff}\{M_i\}_{1\le i \le k}$ denotes for the following affine subspace
$$\text{\rm Aff}\{M_i\}_{1\le i \le k}:=\{M\in \R^p \mid M=\alpha_1M_1+\dots+\alpha_k M_k \text{ \rm where }\alpha_1,\dots,\alpha_p \in \R \text{ \rm and }\alpha_1+\dots +\alpha_p=1\}.$$
The definition of  general position given in Tibshirani \cite{tibshiraniryan} is as follows. 
\begin{defi}[General position]
The $n\times p$ matrix $A=(A_1|\dots|A_p)$ is in general  position if
whatever $s\in \{-1,1\}^p$, whatever  $I\subset \{1,\dots,p\}$ such that ${\rm card}(I)\le \min\{n,p\}$, the affine space $\text{\rm Aff}\{s_iA_i\}_{i\in I}$ satisfies the following property
$$ \forall j \notin I, s_j A_j \notin \text{\rm Aff}\{s_iA_i\}_{i\in I}. $$
\end{defi}
{\bf Example:}
Let $A=(A_1|A_2|A_3)$ and $B=(B_1|B_2|B_3)$ be the following $2\times 3$ matrices
$$A=\begin{pmatrix}1 & 0 & -1/2 \\ 0 & 1 & -1/2\end{pmatrix} \text{ \rm and }B=\begin{pmatrix}1 & 0 & 1 \\ 0 & 1 & 1\end{pmatrix}.$$
Matrix $A$ is not in general position since points $A_1,A_2$ and $-A_3$ are aligned. 
Matrix $B$ is in general position because, up to a sign, $B_1,B_2$ and $B_3$ are not aligned.\\\\ 
Proposition \ref{GPC} shows that when  $X$ is in general position then the  BPDN problem has a unique minimizer.  
\begin{prop}
\label{GPC}
Let $X$ be a $n\times p$ matrix in general position, let $y$ be an arbitrary element of $\R^n$  and let   
$R\ge 0$. Let $S_1^R$ be the solutions of the following optimization problem
$$S^R_1:=\underset{\gamma \in \R^p}{\text{\rm argmin }}\|\gamma\|_1 \text{ \rm subject to }\|y-X\gamma\|^2_2\le R.$$
If $S^R_1$ is not empty then $S^R_1$ is a singleton. 
\end{prop}
When ${\bf 0}$ is an element of $S^R_1$, then $S_1^R=\{{\bf 0}\}$ and thus  proposition \ref{GPC} holds. For this reason, hereafter, it is assumed that ${\bf 0}\notin S_1^R$.  
Proposition  \ref{GPC} is a straightforward implication of Lemmas \ref{famille_libre} and \ref{unicite}.
\begin{lemm}
\label{famille_libre}
Let $X=(X_1|\dots|X_p)$ be a $n\times p$ matrix in general position and let $S_1^R$ as defined in Proposition \ref{GPC}.
If $S^R_1$ is not empty then whatever $\bar{\gamma}\in S_1^R$, the family $(X_i)_{i \in {\rm supp}(\bar{\gamma})}$ is 
linearly independent.
\end{lemm}
{\bf Proof:}
To simplify the notation, without any loss of generality let us assume that ${\rm supp}(\bar{\gamma})=\{1,\dots,k_0\}$.
Let us assume that the family  $(X_i)_{1\le i \le k_0}$ is not linearly independent. What is follows is a contradiction for $X$
in general position. 

Let $b=X\bar{\gamma}$ and let
$\bar{s}=({\rm sign}(\bar{\gamma}_i))_{1\le i \le k_0}$. Because $\bar{\gamma}\in S_1^R$, 
the vector $(\bar{\gamma}_1,\dots,\bar{\gamma}_{k_0})$ is a solution of the following problem 
$$\underset{\gamma\in \R^{k_0}}{\text{\rm argmin }}\underbrace{\sum_{i=1}^{k_0} |\gamma_i|}_{\phi_0(\gamma_1,\dots,\gamma_{k_0})} \text{ \rm subject to the constraint }\forall i \in \{1,\dots,n\}, \underbrace{\sum_{j=1}^{k_0}\gamma_iX_{ij}}_{\phi_i(\gamma_1,\dots,\gamma_{k_0})}=b_i. $$
Because $\phi_0$ is differentiable at $(\bar{\gamma}_1,\dots,\bar{\gamma}_{k_0})$ 
(since $\bar{\gamma}_1\neq 0,\dots,\bar{\gamma}_{k_0}\neq 0$), by the Lagrange multipliers theorem the following implications hold (where $\nabla$ denotes for the gradient).
\begin{eqnarray}
\label{Lagrange}
\nonumber&&\nabla \phi_0(\bar{\gamma}_1,\dots,\bar{\gamma}_{k_0})\in {\rm span}\left\{
\nabla\phi_1(\bar{\gamma}_1,\dots,\bar{\gamma}_{k_0}),\dots,\nabla\phi_n(\bar{\gamma}_1,\dots,\bar{\gamma}_{k_0})\right\},\\
\nonumber&\Rightarrow& \exists \lambda=(\lambda_1,\dots,\lambda_n) \text{ \rm such that }(\bar{s}_1,\dots,\bar{s}_{k_0})=
\lambda_1(X_{11},\dots,X_{1k_0})+\dots+\lambda_n(X_{n1},\dots,X_{nk_0}),\\
&\Rightarrow&\forall i \in \{1,\dots,k_0\}, \bar{s}_i=\lambda_1X_{1i}+\dots+\lambda_nX_{ni}=X_i'\lambda
\end{eqnarray}
Let $(X_i)_{i \in I}$ be a basis of the vectorial space ${\rm span}\left\{X_1,\dots,X_{k_0}\right\}$ (since $(X_i)_{i \in I}$ 
is a basis of a subspace of $\R^n$, then   
${\rm card}(I)\le n$, furthermore ${\rm card}(I)\le k_0\le p$). 
Because the family $(X_i)_{1\le i \le k_0}$ is not then linearly independent, there exists $j\in \{1,\dots,k_0\}\setminus I$
and $(\alpha_i)_{i\in I}$ such that $X_j=\sum_{i\in I}\alpha_iX_i$.
The following implication is a straightforward  consequence of (\ref{Lagrange})
$$\bar{s}_j=X_j'\lambda=\sum_{i\in I}\alpha_iX_i'\lambda=\sum_{i\in I}\alpha_i \bar{s}_i
\Rightarrow 1=(\bar{s}_j)^2=\sum_{i\in I}\alpha_i \bar{s}_i\bar{s}_j.$$
 Finally, the following equality
\begin{equation} \label{barycentre} X_j=\sum_{i\in I}\alpha_iX_i=\sum_{i\in I}\alpha_i(\bar{s}_j\bar{s}_i)^2X_i=\sum_{i\in I}\alpha_i\bar{s}_j\bar{s}_i(\bar{s}_i\bar{s}_jX_i)\end{equation}
 shows that $X_j$ is the barycentre of the family 
$(\bar{s}_i\bar{s}_jX_i)_{i\in I}$ since $1=\sum_{i\in I}\alpha_i \bar{s}_i\bar{s}_j$. Consequently,  equality (\ref{barycentre}) contradicts that $X$ is in general  position. $\hfill \square$

\begin{lemm}
\label{unicite}
Let $S^R_1$ as defined in Proposition \ref{GPC} and let us assume that whatever $\bar{\gamma}\in S^R_1$ the family $(X_i)_{i\in {\rm supp}(\bar{\gamma})}$ is linearly independent then $S^R_1$ is a singleton. 
\end{lemm}
{\bf Proof:}
Let  $\bar{\gamma}^1$ and $\bar{\gamma}^2$ be two elements of $S^R_1$. In the first step, let us show that $X\bar{\gamma}^1=X\bar{\gamma}^2$. 
Since  function $t\in \R^p \mapsto \|y-t\|^2_2$ is strictly convex, if $X\bar{\gamma}^1\neq X\bar{\gamma}^2$ then whatever 
$\alpha \in (0,1)$ the following inequality holds
$$\|y-\alpha X\bar{\gamma}^1 -(1-\alpha)X\bar{\gamma}^2\|_2^2 < \alpha\|y- X\bar{\gamma}^1\|_2^2+(1-\alpha)\|y-X\bar{\gamma}^2\|_2^2\le R.$$
Let us define $\bar{\gamma}_\alpha:=\alpha \bar{\gamma}^1+(1-\alpha) \bar{\gamma}^2$ where $\alpha \in (0,1)$. 
Let us set, for example, $\alpha=1/2$ from the previous inequality. One  may deduce that 1) $\|y-X\bar{\gamma}_{1/2}\|_2^2<R$. Furthermore, 
 since the $l^1$ norm is convex, one may deduce that 2) $\bar{\gamma}_{1/2} \in S^R_1$. Let us show that 1) and 2) provide a contradiction. 
 Since $\|y-X\bar{\gamma}_{1/2}\|_2^2<R$ then there exists $t\in [0,1)$ such that $t\bar{\gamma}_{1/2}$ is
still a feasible point, namely $\|y-Xt\bar{\gamma}_{1/2}\|_2^2\le R$. Because  $\bar{\gamma}_{1/2}\neq {\bf 0}$ (since ${\bf 0}\notin S^R_1$), we have $\|t\bar{\gamma}_{1/2}\|_1<\|\bar{\gamma}_{1/2}\|_1$ which is  contradictory
with $\bar{\gamma}_{1/2}\in S^R_1$. 

To conclude this proof it is sufficient to show that the family $(X_i)_{i \in I}$ where $I={\rm supp}(\bar{\gamma}_1)\cup{\rm supp}(\bar{\gamma}_2)$ is linearly independent. It is straightforward that there exists  $\alpha\in (0,1)$ such that  $I={\rm supp}(\bar{\gamma}_\alpha)$.
Because $X\bar{\gamma}^1=\sum_{i \in I}\bar{\gamma}^1_i X_i=\sum_{i \in I}\bar{\gamma}^2_i X_i=X\bar{\gamma}_2$ and because  the family $(X_i)_{i\in I}$ is linearly independent (since $\bar{\gamma}_\alpha\in S_1^R$) then  
  $\bar{\gamma}_1=\bar{\gamma}_2$. $\hfill \square$\\\\
Proposition \ref{sparsite} shows that once the BPDN (resp. BP) problem has a unique minimizer $\bar{\gamma}$, then 
 the family $(X_i)_{i\in {\rm supp}(\bar{\gamma})}$ is linearly independent. 
\begin{prop}
\label{sparsite}
Let $X$ be a $n\times p$ matrix, let $y$ be an arbitrary element of $\R^n$ and let   
$R\ge 0$. Let us denote $S_1^R$ be the solutions of the following optimization problem
\begin{equation} 
\label{Lasso_contrainte}S^R_1:=\underset{\gamma \in \R^p}{\text{\rm argmin }}\|\gamma\|_1 \text{ \rm subject to }\|y-X\gamma\|^2_2\le R.
\end{equation}
If $S^R_1$ has a unique element $\bar{\gamma}\neq {\bf 0}$, then the family $(X_i)_{i\in {\rm supp}(\bar{\gamma})}$ is linearly independent. 
\end{prop}
Let us assume that  the family $(X_i)_{i\in {\rm supp}(\bar{\gamma})}$ is not linearly independent. We are going to provide a contradiction for the uniqueness of $\bar{\gamma}$ by constructing a feasible point $\tilde{\gamma}$ for which $\tilde{\gamma}\neq \bar{\gamma}$ and $\|\tilde{\gamma}\|_1\le \|\bar{\gamma}\|_1$. Because $(X_i)_{i\in {\rm supp}(\bar{\gamma})}$ is not linearly independent, there exists coefficients $(c_i)_{i\in \supp(\bar{\gamma})}$ not simultaneously null
such that $\sum_{i\in \supp(\bar{\gamma})}c_iX_i={\bf 0}$. This equality provides  an element  $h\in \ker(X)\setminus{\bf 0}$ which is defined hereafter by  
$$\forall i \in {\rm supp}(\bar{\gamma}), h_i=c_i \text{ \rm and } \forall i \notin {\rm supp}(\bar{\gamma}), h_i=0.$$
From this element $h$, one defines feasible points for  problem (\ref{Lasso_contrainte}) by setting for all 
$t\in \R$, $\gamma(t)=\bar{\gamma}+th$ (namely, whatever $t\in \R, \|y-X\gamma(t)\|^2_2\le R$).
Let $f$ be the function $\forall t \in \R, f(t):=\|\gamma(t)\|_1$. 

Now, without loss of generality and to simplify the notation, 
let us assume that ${\rm supp}(h)=\{1,\dots,k_0\}$ thus,
$f(t)=\sum_{i=1}^{k_0}|\bar{\gamma}_i+ th_i|+\sum_{i=k_0+1}^{p}|\bar{\gamma}_i|$.
Whatever $i\in \{1,\dots,k_0\}$, let us denote $t_i=-\bar{\gamma}_i/h_i$ (thus $t_i\neq 0$), the function $t\in \R \mapsto |\bar{\gamma}_i+ th_i|$ is coercive ({\it i.e} $\lim_{t \to +\infty}|\bar{\gamma}_i+ th_i|=\lim_{t \to -\infty}|\bar{\gamma}_i+ th_i|=+\infty$) and affine on the intervals $(-\infty,t_i]$ and $[t_i,\infty)$. 
Consequently, the function $f$ is coercive and affine on the intervals $(-\infty,t_{(1)}],[t_{(1)},t_{(2)}],\dots,[t_{(k_0)},+\infty)$
where $(.)$ is a permutation such that $t_{(1)}\le t_{(2)}\le \dots \le t_{(k_0)}$.
Finally, the minimum of $f$ is reached at $t_0\neq 0$ with  $t_0\in \{t_1,\dots,t_{k_0}\}$, thus there is a feasible point $\tilde{\gamma}=\gamma(t_0)$ such that 
$\tilde{\gamma}\neq \bar{\gamma}$ and $\|\tilde{\gamma}\|_1\le \|\bar{\gamma}\|_1$ which  provides a contradiction. $\hfill \square$
\section*{The stable nullspace property implies the identifiability condition}
Let $X$ be a $n\times p$ matrix, given $\rho\in (0,1)$ the vector $b\in \R^p$ satisfies the stable nullspace property if the following inequality occurs
$$\forall h \in \ker(X), \sum_{i \in {\rm supp}(b)}|h_i|\le \rho \sum_{i \notin {\rm supp}(b)}|h_i|.$$
 One may notice that the stable nullspace property implies the following inequality 
$$\forall h \in \ker(X)\setminus\{0\}, \sum_{i \in {\rm supp}(b)}{\rm sign}(b_i)h_i< \sum_{i \notin {\rm supp}(b)}|h_i|,$$
which characterizes that $b$ is identifiable  with respect to the $L_1$ norm.
Thus the stable nullspace property implies the identifiability condition. 
This statement is not new in the compressed sensing community. Actually, it is well known that the stable nullspace property implies $L_1$ recovery by BP
(see {\it e.g.} \cite{donoho2003,foucartlivre,gribonval}).\\
Now, let us point out an exemple in which the identifiability condition holds but the stable nullspace condition does not hold. 
Let $X$ and $b$ be defined as follows 
$$X:=\begin{pmatrix} 1 & 0 & 1 \\ 0 & 1 & 1 \end{pmatrix} \text{ \rm and }b:= \begin{pmatrix} 0 \\ 1 \\1  \end{pmatrix}.$$
Because $\ker(X)$ is spanned by the vector $(1,1,-1)$, one may deduce that
$$\forall h \in \ker(X), \sum_{i \in {\rm supp}(b)}|h_i|=2 \sum_{i \notin {\rm supp}(b)}|h_i|,$$
thus the stable null space property does not hold.  On the other hand,  
$$\forall h \in \ker(X), \sum_{i \in {\rm supp}(b)}{\rm sign}(b_i)h_i=0,$$
thus the identifiability condition holds.

\bibliographystyle{plain}
\bibliography{bibliothese_active_set}